\documentclass[10pt]{article}

\usepackage{amsmath}
\usepackage{amssymb}

\usepackage{epsf,graphicx}
\include{epsfx}
\usepackage{xtab}
\usepackage{longtable}
\usepackage{cite}

\usepackage[dvips,usenames]{color}

\usepackage[labelfont=bf,labelsep=period,justification=raggedright]{caption}

\makeatletter
\renewcommand{\@biblabel}[1]{\quad#1.}
\makeatother

\date{}

\begin{document}

\begin{flushleft}
{\Large
\textbf{Mechanical strength of 17 134 model proteins
and cysteine slipknots}
}
\\
Mateusz Sikora$^{1,\dagger}$,
Joanna I. Su{\l}kowska$^{1,2,\dagger}$,
Marek Cieplak$^{1,\ast, \dagger}$
\\
\bf{1} Institute of Physics, Polish Academy of Sciences,
Warsaw, Poland
\\
\bf{2} Center for Theoretical Biological Physics, University of California, San Diego,
USA
\\
$\ast$ E-mail:  mc@ifpan.edu.pl
\\
$\dagger$ All authors contributed equally to this work.
\end{flushleft}

\section*{Abstract}
{A new theoretical survey of proteins' resistance to constant speed stretching is performed for a set of 17 134 proteins as described by a structure-based model. The proteins selected have no gaps in their structure determination and consist of not more than 250 amino acids. Our previous studies have dealt with 7510 proteins of no more than 150 amino acids.\\
The proteins are ranked according to the strength of the resistance. Most of the predicted top-strength proteins have not yet been studied experimentally. Architectures and folds which are likely to yield large forces are identified. New types of potent force clamps are discovered. They involve disulphide bridges and, in particular, cysteine slipknots. An effective energy parameter of the model is estimated by comparing the theoretical data on characteristic forces to the corresponding experimental values combined with an extrapolation of the theoretical data to the experimental pulling speeds.\\
These studies provide guidance for future experiments on single molecule manipulation and should lead to selection of proteins for applications. A new class of proteins, involving cystein slipknots, is identified as one that is expected to lead to the strongest force clamps known. This class is characterized through molecular dynamics simulations.}

\section*{Author Summary}
{
The advances in nanotechnology have allowed for manipulation of single
biomolecules and determination of their elastic properties.
Titin  was among the first proteins studied in this way.
Its unravelling by stretching requires a 204 pN force.
The resistance to stretching comes mostly from a localized region
known as a force clamp. In titin, the force clamp is simple as it is
formed by two parallel $\beta-\mathrm{strands}$ that are sheared on pulling.
Studies of a set  of under a hundred of proteins accomplished in the last
decade have revealed a variety of the force clamps that lead to forces
ranging
from under 20 pN to about 500 pN. This set comprises only a tiny fraction
of proteins known. Thus one needs guidance as to what proteins should be
considered for specific mechanical properties. Such a guidance is provided
here through simulations within simplified coarse-grained models on
17 134 proteins that are stretched at constant speed.
We correlate their unravelling forces with two structure
classification schemes. We identify proteins with
large resistance to unravelling and characterize their force clamps.
Quite a few top strength
proteins owe its sturdiness to a new type of the force clamp:
the cystein slipknot in which the force peak is due
to dragging of a piece of the backbone through a closed ring formed by
two other pieces of the backbone and two connecting disulphide bonds.
}

\section*{Introduction}
{
Atomic force microscopy, optical tweezers, and
other tools of nanotechnology have enabled
induction and monitoring of large conformational changes
in biomolecules. Such studies are performed to assess
structure of the biomolecules, their elastic properties, and
ability to act as nanomachines in a cell.
Stretching studies of proteins \cite{encyclopedia} are of a
particular current interest and they have been performed for
under a hundred of systems. Interpretation of some of these experiments
has been helped by all-atom simulations, such as reported in refs.
\cite{Lu1999,P2_t}. They are limited by of order 100 ns time scales
and thus require using unrealistically large constant pulling speeds.
However, they often elucidate the nature of the force clamp --
the region responsible for the largest force of resistance to
pulling, $F_{max}$.
All of the experimental and all-atom simulational studies address merely
a tiny fraction of proteins that are stored in the Protein Data Bank
(PDB) \cite{PDB}. Thus it appears worthwhile to consider a large
set of proteins and determine their $F_{max}$ within an approximate
model that allows for fast and yet reasonably accurate calculations.
Structure-based models of proteins, as pioneered by Go and his
collaborators \cite{Goabe} and used in several implementations
\cite{Veitshans, Hoang1, Clementi, Karanicolas, Hoang3, Hoang2, Hoang, Tozzini},
seem to be suited to this task especially well since
they are defined in terms of the native structures
away from which stretching is imposed.

There are many ways,
all phenomenological, to construct a structure-based model of a
protein. 504 of possible variants are enumerated
and 62 are studied in details in ref. \cite{models}. The variants differ
by the choice of effective potentials, nature of the local backbone
stiffness, energy-related parameters, and of the coarse-grained degrees
of freedom. The most crucial choice relates to making a decision
about which interactions between amino acids count as native contacts.
Comparing $F_{max}$ to the  corresponding experimental values
in 36 available cases selects several optimal models \cite{models}.
Among them, there is one which is very simple and which
describes a protein in terms of its $\mathrm{C}^{\alpha}$ atoms, as labeled
by the sequential index $i$. This model is denoted by
$LJ3=\left\{ \; 6-12,\; C,\; M3,\;{E^0} \;\right\}$ which stands for,
respectively, the Lennard-Jones native contact potentials, local backbone
stiffness represented by harmonic terms that favor the native values of
local chiralities, the contact map in which there are no $i,i+2$ contacts,
and the amplitude of the Lennard-Jones potential, $\epsilon$, is uniform.
The contact map is determined by assigning the van der Waals spheres to the
heavy atoms (enlarged by a factor to account for attraction) and by
checking whether spheres belonging to different
amino acids overlap in the native state \cite{Tsai,Settani}. If they do,
a contact is declared as native. Non-native contacts are considered
repulsive. Application of this criterion frequently selects
the $i,i+2$ contacts as native. If the contact map
includes these contacts the resulting model will be denoted here as
$LJ2$. On average, it performs  worse than $LJ3$ because
the $i,i+2$ contacts usually
correspond to the weak van der Waals couplings as can be demonstrated
in a sample of proteins by using a software \cite{Sobolev} which analyses
atomic configurations from the chemical perspective on molecular bonds.
Thus the $i,i+2$ couplings should better be removed from the contact map
(in most cases).

The survey to determine $F_{max}$ in
7510 model proteins with the number of amino acids, $N$, not exceeding
150 and 239 longer proteins (with $N$ up to 851) has been accomplished
twice. First within the $LJ2$ model \cite{JPCM} and soon afterwords
within the $LJ3$ model \cite{BJ}.
The first survey also comes with many details
of the methodology whereas the second just presents the outcomes.
The two surveys are compared in more details in refs. \cite{models,acta}.
The results differ, particularly when it comes to ranking
of the proteins according to the value of $F_{max}$, but they
mutually provide the error bars on the findings. They both agree,
however, on predicting that there are many proteins whose strength
should be considerably larger than the frequently studied benchmark --
the sarcomere protein titin ($F_{max}$ of order 204 pN \cite{R4,C6}). Near
the top of the list, there is the scaffoldin protein c7A (the PDB code 1aoh)
which has been recently measured to have $F_{max}$ of about
480 pN \cite{Valbuena}. Other findings include establishing correlations
with the CATH hierarchical
classification scheme \cite{Orengo1997,Pearl2005}, such as that there are no
strong $\alpha$ proteins, and identification of several types of the
force clamps. The large forces most commonly originate in
parallel $\beta-\mathrm{strands}$ that are sheared \cite{B2}. However, there
are also clamps with antiparallel $\beta-\mathrm{strands}$, unstructured strands,
and other kinds.

The two surveys have been based on the structure download  made on July 26, 2005
when the PDB comprised 29 385 entries. Many of them correspond to
nucleic acids, complexes with nucleic acids and with other proteins,
carbohydrates, or come with
incomplete files and hence the much smaller number
of proteins that could be used in the molecular dynamics studies.
Here, we present results of still another survey which is based on a
download of December 18, 2008  which contains 54 807 structure files
and leads to 17 134 acceptable structures with $N$ not
exceeding 250 (instead of 150).
These structures are then analyzed through simulations based on the
$LJ3$ model. The numerical code has been improved
to allow for acceleration of calculations by a factor of 2.

The 190 structures (or 1.1 \% of all structure considered)
with the top values of $F_{max}$ in units of
$\epsilon/\mathrm{\AA}$ are shown in Table 1  (the first 81 entries for which
$F_{max} \ge 3.9\: \epsilon/\mathrm{\AA}$) and Table S1 of the SI
(proteins ranked 82 through 190), together with the values of titin
(1tit) and ubiquitin (1ubq) to provide a scale. As argued in
the Materials and Methods section
section, the unit of force, $\epsilon/\mathrm{\AA}$,
is now estimated to be of order 110 pN.
All of the corresponding proteins are predicted to be much stronger
than titin and none but two of them (1aho, 1g1k \cite{Valbuena})
have been studied experimentally yet.
In addition to the types of force clamps identified before, we have
discovered two new mechanisms of sturdiness.
One of them involves a cysteine slipknot (CSK) and is found to be
operational in all of the 13 top strength proteins.
In this motif, a slip-loop is pulled out of a cysteine knot-loop.
Another involves dragging of a single fragment of the main chain
across a cysteine knot-loop.
The two mechanisms are similar in spirit since both involve
dragging of the backbone. However, in the CSK case, two fragments
of the backbone are participating.


We make a more systematic identification of the CATH-classified
architectures that are linked to mechanical strength and then
analyze correlations of the data to the SCOP-based
grouping (version 1.73) \cite{Murzin1995,Andreeva2008,Lo_Conte2002}.
The previous surveys did not relate to the SCOP scheme.

We identify the CATH-based architectures and SCOP-based folds
that are associated with the occurrence of a strong resistance
to pulling. A general observation, however, is that
each such group of structures may also include
examples of proteins that unravel easily.
The dynamics of a protein are very sensitive to
mechanical details that are
largely captured by the contact map and not just by the
appearance of a structure.
On the other hand, if one were to look for mechanically
strong proteins then the architectures and folds identified
by us should provide a good starting point.
We also study the dependence of $F_{max}$ on the pulling velocity
and characterize the dependence on $N$ through distributions of
the forces.

The current third survey has been performed within the same $LJ3$ model
as the second survey \cite{BJ}. However, we reuse and extend it here
because the editors of Biophysical Journal retracted the second survey
\cite{retraction}. All of the values of $F_{max}$ are deposited at the
website www.ifpan.edu.pl/BSDB (Biomolecule Stretching Database) and
can by accessed by through the PDB structure code.

\section*{Results and Discussion}
{

\subsection*{Distribution of Forces}

The distribution of all values of $F_{max}$ for the full set
of proteins is shown in Figure \ref{allhist}.
Despite the larger limit on $N$ now allowed, the distribution is
rather similar to that obtained in ref. \cite{BJ} for the smaller number of
proteins (and with the smaller sizes). The similarity is primarily
due to the fact that the size related effects, discussed below,
are countered by new types of proteins that are now incorporated
into the survey.
The distribution is peaked around $F_{max}$ of $1.2\: \epsilon/\mathrm{\AA}$
which constitutes about 60\% of the strength associated with titin.
The distribution is non-Gaussian: it has a zero-force peak and a long
force tail. The zero-force peak arises in some proteins with the covalent
disulphide bonds. In the model, such bonds are represented
by strong harmonic bonds. Stretching of such a protein may not result in any
force peak before a disulphide bond gets stretched indefinitely
and hence $F_{max}$ is considered to be vanishing then.
The tail, on the other hand, corresponds to the strong proteins.
The top strongest 1.1\% of all proteins are listed in Tables 1
(in the main text) and S1 (in the SI).

The insets of Figure \ref{allhist} show similar distributions for
proteins belonging to the particular CATH-based classes.
There are four such classes:
$\alpha$, $\beta$, $\alpha- \beta$ and proteins with no apparent
secondary structures.
It is seen that none of the 3240 $\alpha$ proteins exceeds the peak force
obtained for titin within our model.
This observation is in agreement with experiments
on several $\alpha$ proteins that are listed in the Materials and Methods section.
All strong proteins are seen to involve the $\beta-\mathrm{strands}$.
The peak in the probability distribution
for the $\alpha-\beta$ proteins is observed to be shifted towards the
bigger values of $F_{max}$ compared to the one for the $\beta$ proteins.
At the same time,  the high force tail of the distribution for the $\beta$
proteins is substantially more populated than the corresponding tail for
the $\alpha$ proteins.

Figure \ref{sizedep} is similar to Figure \ref{allhist} in spirit, but
now the structures are split into
particular ranges of the protein  sizes: $N$ between
40 and 100 (the dotted line), between 100 and 150 (thin solid line),
and between 200 and 250 (the thick solid line). The curve for the range
from 150 to 200 is in-between the curves corresponding to neighboring
ranges and is not shown in order not to crowd the Figure. The distributions
are seen to be shifting to the right when increasing the range of the values
of $N$ indicating, that the bigger the number of amino acids, the more likely
a protein is to have a large value of $F_{max}$. This observation holds for
all classes of the proteins, as evidenced by the insets in
Figure \ref{sizedep}.

In most cases, the major force peak arises at the begining of stretching
where the Go-like model should be applicable most adequately.
One can characterize the location of $F_{max}$ during the
stretching process by a dimensionless parameter $\lambda$ which
is defined in terms of the end-to-end distance, as spelled out
in the caption of Table 1. This parameter is equal to 0 in the native
state and to 1 in the fully extended state.
In 25 \% of the proteins studied in this survey, $\lambda$ was
less than 0.25 and in 52 \% -- les than 0.5.
There are very few proteins with $\lambda$ exceeding 0.8.

Table 1 does not include any (non-cysteine-based) knotted proteins. The full
list of 17 134 proteins contains 42 such proteins but they come with moderate
values of $F_{max}$. However, knotted proteins with $N > 250$ may turn out
to have different properties.

\vspace*{0.5cm}
\subsection*{Biological properties of the strongest proteins}

A convenient way to learn about the biological properties listed in Tables 1 and S1
is through the Gene Ontology data base \cite{GO} which links such properties
with the PDB structure codes. The properties are divided into three domains.
The first of these is "molecular function" which describes a molecular function
of a gene product. The second is "biological processes" and it covers sets of
molecular events that have well defined initial and final stages. The third is
"cellular component" and it specifies a place where a given gene product is
most likely to act.

The results of our findings are summarised in Table 2.
It can be seen, that most of the 190 strongest proteins are likely
to be found in an extracellular space where conditions are much more
reducing than within cells. Larger mechanical stability is advantageous under
such conditions. 90 out of the strongest proteins exhibit hydrolase activity.
39 of these 90 are serine-type endopeptidases. These findings
seem to be consistent with expectations regarding proteins endowed with
high mechanical stability. For instance, proteases, which are well represented
in Table 2 should be more stable to prevent self-cleavage.

\vspace*{0.5cm}

\subsection*{CATH-based architectures}

The classification of proteins within the CATH (Class, Architecture, Topology,
Homology) data base is done
semi-automatically by applying numerical algorithms  to
structures that are resolved better than within 4 {\AA}
\cite{Orengo1997,Pearl2005}.
The four classes of proteins in the CATH system
are split into architectures, depending on the overall
spatial arrangement of the secondary structures, the numbers
of $\beta-\mathrm{sheets}$ in various motifs, and the like. The next finer
step in this hierarchical scheme is into topologies and it involves
counting contacts between amino acids which are sequentially
separated by more than a treshold.
The further divisions into homologous superfamilies and then sequence
family levels involve studies of the sequential identity.

We have found that  only six architectures contribute to $F_{max}$ larger
than $4\: \epsilon/\mathrm{\AA}$.
These are ribbons -- 2.10 (41.8 \% of the proteins listed in Table 1),
$\beta-\mathrm{barrels}$ -- 2.40 (8.9 \%), $\beta-\mathrm{sandwiches}$ -- 2.60 (16.3 \%),
$\beta-\mathrm{rolls}$ -- 3.10 (5.4 \%), 3-layer (aba) sandwiches -- 3.40 (5.4 \%),
and these with no CATH classification to date (21.8 \%).
The corresponding distributions of forces are shown in the top six panels of
Figure \ref{catharch} and the topologies involved
are listed and named in Table 3.

Examples of architectures that are dominant contributors to
a low force behavior are the $\alpha$ orthogonal bundle (the right
bottom panel of Figure \ref{catharch}), the $\alpha$ up-down bundle,
and the $\beta-\mathrm{roll}$ (the left bottom panel of Figure \ref{catharch}).

\vspace*{0.5cm}

\subsection*{ SCOP-based classes and folds}

The SCOP (Structural Classification of Proteins)
data base \cite{Murzin1995,Andreeva2008,Lo_Conte2002} is curated
manually and it relies on making comparisons to other
structures through a visual inspection. This classification scheme
is also hierarchical and the broadest division
is into seven classes and three quasi-classes.
The classes are labelled $a$ through $g$ and these are as follows:
mainly $\alpha$ ($a$), mainly $\beta$ ($b$), $\alpha/\beta$ which groups
proteins in which helices and $\beta-\mathrm{sheets}$ are interlaced ($c$),
$\alpha + \beta$ with the helices and $\beta-\mathrm{sheets}$ grouped into
clusters that are separated spatially ($d$), multidomain proteins ($e$),
membrane and cell-surface proteins ($f$), and small proteins that are
dominated by disulphide bridges or the heme metal ligands ($g$).
The quasi-classes are labelled $h$ through $j$ and they comprise
coiled-coil proteins ($h$), structures with low resolution ($i$), and
peptides and short fragments ($j$). The classes are then partitioned
into folds that share spatial arrangement of secondary structures
and the nature of their topological interlinking. Folds are then divided
into superfamilies  (same fold but small sequence identity) and then
families (two proteins are said to belong to the same family if
their sequence identity is at least 30\%). Families are then
divided into proteins -- a category that groups similar structures
that are linked to a similar function. Proteins comprise various
protein species.

Each structure assignment comes with an alphanumeric label, as shown
in Tables 1, S1, and 4 which reflects the placement in the hierarchy.
At the time of our download, there have been 92 972 entries in the SCOP
data base that are assigned to 34 495 PDB structures.
These entries are divided into 3464 families,
1777 superfamilies and 1086 unique folds.
A given structure may have several entry labels but the dominant
assignment is listed first. We use the primary assignment in our studies.
The same rule is also applied to the CATH-based codes.

Figure \ref{scop} shows the distributions of forces for the
SCOP-based classes of proteins. The results are consistent with
the CATH-based classes since the $\alpha-\beta$ class of CATH
basically encompasses the $\alpha/\beta$ and $\alpha+\beta$ classes of SCOP.
However, there are proteins which are classified only according
to one of the two schemes. Thus there are 4431 $\alpha-\beta$
proteins out of which only the total
of 3368 is SCOP-classified as belonging to the $\alpha+\beta$
and $\alpha/\beta$ classes. At the same time, the total of the
proteins in the $\alpha+\beta$ and $\alpha/\beta$ classes we
have is 4795.

It should be noted that the peak in the distribution for
$\alpha+\beta$ is shifted to higher forces by about $0.7\:
\epsilon/\mathrm{\AA}$ from the peak for $\alpha/\beta$. At the same time,
the zero-force peak is virtually absent in $\alpha+\beta$.
The SCOP-based classification also reveals that its class $g$ contributes
across the full range of forces and, in particular, it may lead to
large values of $F_{max}$. It should be noted, as also evidenced by
Table 1, that there is a substantial number of strong proteins
that has no class assignment.

Figures \ref{scop1} and \ref{scop2} refer to the distributions of
$F_{max}$ across specific folds. The first of these presents
results for the folds that give rise to the largest forces. The names
of such folds are specified in Figure \ref{scop1}.
The percentage-wise assessment of the folds contributing to
big forces is presented in Table 4. The top contributor is found
to be the $b.47$ fold (SMAD/FHA domain).
Figure \ref{scop2} gives examples of folds that typically yield low forces.

It is interesting to note that distributions corresponding to some folds
are distinctively bimodal, as in the case of the SMAD/FHA fold (b.47).
This particular fold is dominated by SMAD3 MH2 domain (b.47.1.2; 352 structures)
which contributes both to the high and low force peaks in the distribution.
The remaining domains (b.47.1.1, b47.1.3, and b47.1.4) contribute only
to the low force peak. The dynamical bimodality of the b.47.1.2 fold can be
ascribed to the fact that the strong subset comes with one extra disulphide bond
relative to the weak subset. This extra bond provides substantial additional
mechanical stability when stretching is accomplished by the termini. We illustrate
sources of this bimodality in the SI (Figure S1) for two proteins from this fold:
1bra which is strong and 1elc which is weak. In ref. \cite{JPCM}, we have noted
that various sets of proteins with identical CATH codes (e.g., 3.10.10) may give
rise to bimodal distributions without any dynamical involvement of the
disulphide bonds. The reason for this is that even though the contact maps for the
two modes are similar, the weaker subset misses certain longer ranged contacts which
pin the structure. Mechanical stability is more sensitive to structural and dynamical
details than are not provided by standard structural descriptors.

\vspace*{0.5cm}

\subsection*{Force clamps}

{\bf Shearing motif}.
The most common type of the force clamp identified in the literature
is illustrated in the
top left panel of Figure \ref{cysclamp} corresponding to the 14th-ranked
protein 1c4p. In this case, the strong resistance to pulling is due to
a simultaneous shearing of two $\beta-\mathrm{strands}$ which are additionally
immobilised by short $\beta-\mathrm{strands}$ that adhere to the two strands.
Similar motifs appears in 1qqr(15), 1j8s(17), 1j8r(19), 1f3y(20), 2pbt(29),
2fzl(15), 1aoh(19), where the number in brackets indicate ranking as shown in Table 1.
It is interesting to note that the $\beta-\mathrm{strands}$ responsible for the mechanical
clamp in 1j8s and 1j8r display an additional twist. Undoing the twist enhances
$F_{max}$. (There is a similar mechanism that seems to be
operational in the case of a horseshoe conformation
 found in ankyrin \cite{L4,ankyrin}).
The force clamps are identified by investigating the effect of removal of
various groups of contacts on the value of $F_{max}$ \cite{Hoang,JPCM}.

There are, however, new types of the force clamps that we observe in the
proteins listed in Tables 1 and S1. They arise from entanglements resulting
from the presence of the disulphide bonds which cannot be ruptured by forces
accessible in the atomic force microscopy.
We note that about 2/3 of the proteins listed in Tables 1 and S1
contain the disulphide bonds. Many of these bonds do not carry
much of dynamical relevance  when pulling by the termini.
However, in certain situations they are the essence of the force clamp.
The disulphide bonds have been already identified as leading to
formation of the cystein knot (CK) motifs \cite{Craik,Gruber}
(such proteins are found in the toxins of spiders and scorpions)
and the cyclic CK motifs \cite{Craik1,Craik2}.
Here, we find still another motif -- that of the CSK which is
similar to that found in slipknotted proteins \cite{Yeates_2007b,Taylor_2007,Sulkowska_2009}
which do not conatin the disulphide bonds. This motif is found in the top 13 proteins.
The cysteine loop,  knot,  and slipknot motifs are shown schematically
in the remaining panels of Figure \ref{cysclamp}. It is convenient to divide
these motifs into two categories:
shallow (S) and deep (D) (according to the classification used for knotted proteins
\cite{Virnau,Sulkowska_2008e}), depending on whether the motif is
spanning most of the sequence or is instead localized in its small
fraction.

{\bf Shearing connected with a cysteine loop}.
In this case, the mechanical clamp arises from shearing between a $\beta-\mathrm{strand}$
belonging to a deep cysteine loop and another
strand located outside the loop (the left bottom panel of Figure \ref{cysclamp}).
Existence of the disulphide bond before the shearing motif
allows to decompose direct tension onto the $\beta-\mathrm{strands}$
making the protein resist stretching much more effectively
than what would be expected from a simple shearing motif.
Additionally, the disulphide bonds prevent an onset of any
rotation in the protein conformation which otherwise might form
an opportunity for unzipping.
This motif appears in 1dzj(40,D)
1vsc(37,D), 1dzk(35,D), 1i04 (81,D), 1hqp(83,D),
1oxm(98,D),  2a2g (175,D), 2boc(179,D), and many other proteins.
The middle panel of Figure \ref{deep} gives an example of the
corresponding force ($F$) -- displacement ($d$) pattern as obtained
for 1dzj.

{\bf Shearing and dragging out of a cysteine loop}.
This motif consists of two parts.
The first is formed by a rather small and deep cysteine loop
which is located very close to
one terminus with the second terminus located across the cysteine loop.
The motif arises when almost all of the protein backbone is dragged
across the cysteine loop on stretching. A protein structure
also contains a few $\beta-\mathrm{strands}$ which get sheared before dragging takes
place. This motif is seen in 1kdm(23,D), 1q56(24,D), 1qu0(33,D), 1f5f(34,D)
and this geometry of pulling we call geometry I.
It should be pointed out that, in all such cases,
pulling by the N terminus takes place within
(or very near) the plane formed by the cysteine loop.
A small change in such a geometry,
e.g. the one arising from pulling not by the last amino acid
but by the penultimate bead, may cause getting out of
the cystein loop and result in a very
different unfolding pathway with a distinctly different value of $F_{max}$.
In this other kind of pulling set up, denoted as geometry II, the loop is
bypassed and the resistance to pulling is provided only
by the shearing mechanism.

Dragging arises from overcoming steric constraints and generates
an additional contribution
to the strength of the standard shearing mechanical clamp.
By using geometry II and also by eliminating
the native contacts between the sheared $\beta-\mathrm{strands}$ we can
estimate the topological contribution
of the dragging effect on the value of $F_{max}$.
For proteins 1kdm, 1q56, 1qu0, 1f5f, it comes out to be around 25 \%.
The force $F$--$d$ patterns corresponding to these
two geometries of pulling are shown
in top panel of Figure \ref{pierscien1}.

In the survey, there are other proteins which also have disulphide bonds and
belong to the 2.60.120.200 category. These proteins have a cysteine which
is either very shallow or deep, but is located in the middle of the
protein backbone so that there is no possibility to form a long $\beta-\mathrm{strand}$.
In this case, the dragging effects are much smaller. For instance,
for 1pz7(D) and 1cpm(S), $F_{max}$ is close to $1\: \epsilon/\mathrm{\AA}$.

{\bf Shearing inside of a cysteine knot }.
This motif is created by a loosely
packed CK
(two or more spliced cysteine loops) with at least two parallel $\beta$
strands that are present within the knot.
Pulling protein by termini exerts tension on the entire CK
and thus produces an indirect shearing force on the $\beta-\mathrm{strands}$
inside the entangled part of the protein.
In this case, elimination of the native contacts between the
$\beta-\mathrm{strands}$ reduces
$F_{max}$ only partially indicating that the mechanical clamp is created also
by the CK.
A simple CK is also found  in 2bzm(42) and many other proteins, e.g.
in 2g7i(77,S), 1hfh103,S), 2g4x(136,D), 2g4w(169,D).
The $F-d$ patterns for 2bzm and 2g4x are shown in the bottom
panel of Figure \ref{deep}.
More complex structures or higher order CKs
(with more than two cystein bonds)
can be identified in 1afk(85), 1afl(117), or 1aqp(135).
Inside this group of proteins there are also examples of
proteins -- 1qoz(88,S) --
in which a cysteine loop is braided
to a CK by some native contacts.

{\bf Cysteine slipknot force-clamp}  is observed in the strongest 13 proteins.
The top strength protein is 1bmp (bone morphogenic protein)
with the predicted $F_{max}$ of $10.2\: \epsilon/\mathrm{\AA}$, which should
correspond to about 1100 pN (see Materials nad Methods). This strength
should be accessible to standard experiments as the atomic force microscopy
has been already used to rupture covalent N-C and C-C bonds
by forces of 1500 and 4500 pN respectively \cite{grandbois}.

In our discussion, we focus on the 13-ranked 1vpf (a vascular endothelial
growth factor) with the predicted $F_{max}$ of $5.3\: \epsilon/\mathrm{\AA}$.
The CSK motif arises from two loops \cite{Sulkowska_2009}:
the knot-loop and the slip-loop, where the slip-loop can be
threaded across the knot-loop. One needs at least three
disulphide bonds for this motif to arise.

In the case of the 1vpf, the knot-loop is created by the
disulphide bonds between amino acids
57 and 102, 61 and 104, and the protein backbone between amino acids
57-61 (GLY,GLY,CYS) and 102-104 (GLU).
The slip-loop is created by the protein backbone between
sites 61-102 and is stabilized
by  12 hydrogen bonds between two parallel $\beta-\mathrm{strands}$.
In the CSK motif, the force peak is due to dragging of a slip-
loop through the knot-loop making the native hydrogen contacts only
marginally responsible for the mechanical resistance. Thus the force peak
arises, to a large extent, from overcoming steric constraints,
i.e. it is due to repulsion resulting from the excluded volume.
The $F-d$ pattern for this novel type of a force clamp is shown in
the top panel of Figure \ref{deep}. Another example of such a
pattern for a CSK is shown
in the bottom panel of Figure \ref{pierscien1} for the 22nd ranked 2h64
(a human transforming growth factor).
The leading role of the steric constraints is verified by checking the
reduction of the $F_{max}$ when all the slipknot-related contacts
(inside the slip-loop and between the slip-loop and the knot-loop)
are converted to be purely repulsive. As a result of this bond removal,
the force peak persists, though it gets shifted and becomes smaller.
This is summarized in Table S2 in the SI. It is a new and unexpected result.

Another way to establish the role of the CSK motif
is to create the disulphide-deficient mutants, as accomplished
experimentally \cite{Welfle} for 1vpf.
The two mutants, 1mkk (C61A and C104A)
and 1mkg (C57A and C102A), have structures similar to 1vpf
but contain no knot-loops and thus there is no slipknot.
Muller et al. \cite{Welfle} show that the mutants' thermodynamic
stability is not reduced
but their folding capacity is.
Our work shows that the mutants have a reduced resistance to pulling
compared to 1vpf: $F_{max}$ drops from $5.3\: \epsilon/\mathrm{\AA}$
to 1.49 and 2.01 $\epsilon/\mathrm{\AA}$ for 1mkk and 1mkg respectively.

We note that the CSK topology is a subgroup inside the
CK class (represented mostly by 2.10.90.10) and the CSK force clamp
need arise for a particular way of pulling. For instance,
proteins 1afk(68), 1afl(100) or 1aqp(118) have up to four
disulphide bonds and yet the CSK
motif does not play any dynamical role in pulling by the
terminal amino acids. In the case of the CSK,
we observe a formidable dispersion in the values of
$F_{max}$. For example, it ranges  between 4.8--5.9, 4.1--4.8, and
4.1--5.2 $\epsilon/\mathrm{\AA}$  for various trajectories in
1vpf, 2h64, and 2c7w respectively. We now examine the CSK geometry
in more details.

{\bf Cysteine slipknot motif} is distinct from the slipknot motif in
several ways. The left-most panel of Figure \ref{cskloop}
shows a slipknot with three intersections at sequential locations
$k_1$, $k_2$, and $k_3$. This geometry is topologically trivial since
when one pulls by the termini, the apparent entanglement may untie and
become a simple line. The entanglement would form the trefoil knot if the
$k_3$ intersection was removed by redirecting the corresponding segment
of the chain (thin line) away from the $k_1 - k_3$ loop.
Such slipknot motifs have been observed
in native states of several proteins \cite{Yeates_2007b,Taylor_2007,Sulkowska_2009}.
In contrast, the CSKs are not present in the native state but arise as a result
of mechanical manipulation. The middle panel of Figure \ref{cskloop} shows a schematic
representation of a native conformation with three cysteine bonds:
between $i_1$ and $j_1$, between $i_2$ and $j_2$, and between $i_3$ and $j_3$.
The $i$-ends of the bonds are counted as being closer to the N-terminus.
The three bonds are in a specific arrangement as shown in the panel.
In particular, the $i_3-j_3$
bond must cross the loop $i_1-i_2-j_2-j_1$. This loop consists of two pieces of the
backbone ($i_1 - i_2$ and $j_2 - j_1$) that are linked to form a closed path
by the two remaining cysteine bonds -- it is the cysteine knot-loop.
The average radius of this loop is denoted by $R_{ck}$.

The arrangement shown in the middle panel has no entanglements that could
be considered as knots in the topolgical sense.
However, on pulling by the termini,
the chain segment adjacent to $i_3$ gets threaded through the knot-loop
since $i_3$ is rigidly attached to $j_3$, as illustrated
in the rightmost panel of Figure \ref{cskloop}. Pulling by
$i_3-j_3$ also results in generating another loop -- the cysteine slip-loop --
since the segment around $i_3$ gets bent strongly to form a cigar like shape
with the radius of curvature at the $i_3$-tip denoted by $R_{cs}$.
This loop extends between $i_2$ and $j_1$.
It should be pointed out that the cysteine knot-loop in the CSK is stiff
whereas in a slipknotted protein (such as the thymine kinase)
its size is variable (as it can be tightened on the protein backbone
\cite{Sulkowska_2009}
in analogy to tightening a knot \cite{Sulkowska_2008} by pulling).

The dynamics of pulling depends of the relationship between $R_{ck}$ and $R_{cs}$
as the ''cigar" may either go through or get stuck. In the former case a related
force peak would arise. If the system was a homogeneous polymer, dragging
would be successful when $R_{ck}$ was bigger than $R_{cs}$. The corresponding
force would be related to the work against the elasticity that was needed to bend
the slip-loop to the appropriate curvature. This work is proportional to the
square of the curvature. Thus the total elastic energy involved in bending
the segment $i_2-j_1$ is of order
 $\oint ds R^{-2} \sim R^{-1}_{cs}$ \cite{Landau}, where $s$ is the arc distance.
Dividing this energy by the distance of pulling would yield an estimate of the force
measured if thermal fluctuations were neglected.
The geometrical condition for dragging in proteins is more complicated because
of the presence of the side groups and the related non-homogeneities and variability
across the hydrophobicity scale. The diameter of the ''rope" that the knot
loop is made of should not exceed the maximum a linear extension, $t_k$ of amino
acids. Thus the effective inner radius of the knot-loop is $R_{ck} - t_k$.
Similarly, the size of the outer circle that is tangential to the tightest slip-loop
is $R_{cs} + t_s$, where $t_s$ is the thickness of the slip-loop.
(Both thicknesses can be considered as being site dependent and including
possible hydration layer effects near polar amino acids.)
Thus the slip-knot can be driven through the cystein knot-loop provided
\begin{equation}
R_{cs} + t_s  \; < \;R_{ck} - t_k  \;\;.
\end{equation}
In our simulations, the successful threading situations correspond to
$R_{ck}$ and $R_{cs}$ of around 7 and 3 {\AA}. The amino acids in the knot-loop
are mostly Gly, Ala, or Cys with their side groups pointing outside of the
loop. One may then estimate $t_k$ to be about 1.5 {\AA}. On the other hand,
the linear size of the amino acids in the slip-loop can be determined to be
close to 2.5 {\AA}. These estimates indicate that $R_{cs} + t_s$ can be very close
to $R_{ck} - t_k$ so the possibility of slipping through the knot-loop
is borderline. In fact, slipping might be forbidden within the framework
of the tube-picture of proteins \cite{tube,tube1} in which the effective
thickness of the tube is considered to be 2.7 {\AA}.

The CSK motifs give rise to a force peak in 1vpf, 2h64(22,S), 1rv6(25,S), 1waq(26,S),
1reu(27,S), 1tgj(28), 2h62(30,S), 1tgk(31), 2c7w(38,D), 2gyr(39,S), 1lx5(95,D),
and many other proteins. In these cases, the typical value of $R_{ck}$ is  about 7 {\AA}.
However, specificity may result in somewhat smaller values of $R_{ck}$
which may cause only smaller segments of the slip-loop to be threaded.
If the passage is blocked, there will be no isolated force peak
as happens in 1tgj and 1vpp.

{\bf Types of the force--displacement patterns for proteins with the disulphide bonds}.
In the case of proteins with very shallow cystein knot, loop or slipknot
motifs, $F$ increases very rapidly with $d$ and isolated force peak does
not arise ($F_{max}=0$).
Such cases are represented, e.g., by 1bmp, 1rnr, 1ld5, and 1wzn
where the slipknots are either very tight or the cystein
loop is very shallow.
In the case of a shallow motif, however, a force peak
can sometimes be isolated as in the case of the 13th-ranked
protein 1vpf (Figure \ref{deep}) and in several other proteins,
like 1xzg and 1dzk.
In this case, the value of $F_{max}$ takes into account
tension on the cystein bonds and it is not obvious whether
such a strong elastic background should be subtracted from
the value of $F$ when determining $F_{max}$ or not.
In this survey, we do not subtract the backgrounds.
It should be noted that in our previous surveys we
missed the CSK-related force peaks
because we attributed the rapid force rises at the end of
pulling just to stretching of the backbone without
realizing existence of structure in some such rises.

For a deep motif, the $F-d$ pattern may have several
small force peaks
before the final rise of the force, as observed for 2g4s and 1bj7.
When the CSK motif is very deep, it usually does not
have any influence
on the shape of the $F-d$ pattern apart from a much steeper final
rising force. Such a situation is seen in the case of, e.g., 1j8r and 1j8s.



\vspace*{0.5cm}
\subsection*{Concluding remarks}

This surveys identifies a host of proteins that are likely
to be sturdy mechanically. Many of them involve disulphide
bridges which bring about entanglements that are complicated
topologically such as CSKs and CKs.
The distinction between the two is that the former can depart
from its native conformation and the latter cannot.

Our survey made use of a coarse grained model
so it would be interesting to reinvestigate some of the
proteins identified here by all-atom simulations, especially
in situations when the CSK is involved.
The CSK motifs may reveal different mechanical properties
when studied in a more realistic model. Of course, a
decisive judgment should be provided by experiment.

The very high mechanical resistance of the CSK proteins
should help one to understand their biological function.
The superfamily of cysteine-knot cytokines (in class small proteins
and fold cystein-knot cytokines) includes families of
the transfroming growth factor (TGF)--$\beta$ and the polypeptide
vascular endothelial growth factors (VEGFs) \cite{Iyer,Well}.
The various members of this superfamily, listed in Table 5,
have distinct biological functions. For instance,
VEGF-B proteins which regulate the blood vessel and
limphatic angiogenesis bind only to one receptor of tyrosine kinase VEGFR-1.
On the other hand, VEGF-A proteins bind to two receptors
VEGFR-1 and VEGFR-2. All of these proteins form a dimer structure.
The members of this familly are endowed with
remarkably similar monomer structures
but differ in their mode of dimerisation and thus
in their propensity to bind ligands.
Additionally, all dimers posses almost the same a cyclic
arrangement of cysteine residues which 
are involved in both intra- and inter-chain disulphide bonds.
These inter-chain disulphide bonds create the knot and
slip-loops, where the intra-chain
disulphide bonds  give rise to a CSK motif when
the slip-loop is gets dragged acrros the knot-loop upon pulling.

It has been shown experimentally \cite{Choe} that such cysteine related connectivities
bring the key residues involved in receptor recognition
into close proximity of each other.
They also provide a primary source of stability
of the monomers due to the lack of other
hydrogen bonds between two beta strands at the dimer interface.

The non trvial topologial connection between the monomers allow for mechanical
separation of two monomers by a distance of about half of the size of the slip-loop.
Our results suggest, however, that the force needed for the separation may be
too high to arise in the cell.

}


\section*{Materials and Methods}
{
The input to the dynamical modeling is provided by a PDB-based structures.
The structure files may often contain several chains. In this case,
we consider only the first chain that is present in the PDB file.
Likewise, the first NMR determined structure is considered.
If a protein consists of several domains, we consider only the first of them.

The modeling cannot be accomplished if a structure
has regions or strings of residues which are not sufficiently
resolved experimentally. Essentially all structure-disjoint
proteins have been excluded for our studies.
Exceptions were made for the experimentally studied scaffoldin 1aoh
and for proteins in which small defects in the established structure
(such as missing side groups)
were confined within cystein loops and were thus irrelevant dynamically.
In these situations, the missing contacts
have been added by a distance based criterion \cite{Valbuena}
in which the treshold was set at 7.5 {\AA}.
Among the test used to weed out inadequate structures involved
determining distances between the consecutive
$\mathrm{C}^{\alpha}$ atoms. A structure was rejected if these distances
were found to be outside of the range of 3.6-3.95 \AA.
The exception was made for prolines, which in its native state can
accommodate the cis conformation. In that case, the distance between
a proline $\mathrm{C}^{\alpha}$ and its subsequent amino acid
usually  falls in the range between 2.8 and 3.85 {\AA}.
For a small group of proteins which slipped through our structure
quality checking procedure, but were found to be easily fixed
(e.g. 1f5f, 1fy8, and 2f3c),
we used publicly avialable software  BBQ \cite{Gront_2007}
to rebuild locations of the missing residues.
A limited accuracy of this prediction procedure seems
to be adequate for our model due to its the coarse-grained nature.

The modeling of dynamics follows our previous implementations
\cite{Hoang2,Hoang,JPCM} within model $LJ2$ except that the
contact map is as in ref. \cite{BJ}, i.e. with the $i,i+2$ contacts excluded.
There is also a difference in description of the disulphide bonds.
In refs. \cite{BJ,models} they were treated as an order-of-magnitude
enhancement of the Lennard-Jones contacts in all proteins.
In ref. \cite{JPCM}
the different treatment of the disulphide bonds was applied to the
proteins that were found to be strong mechanically without any
enhancements.
Here, on the other hand, we consider such bonds as harmonic
in all proteins, in analogy to the backbone links between
the consecutive $\mathrm{C}^{\alpha}$s.
The native contacts are described by the Lennard-Jones potential
$V^{6-12}=4\;\epsilon
[\left(\frac{\sigma_{ij}}{r_{ij}}\right)^{12}-
\left(\frac{\sigma_{ij}}{r_{ij}}\right)^{6}]$, where
$r_{ij}$ is the distance between the $\mathrm{C}^{\alpha}$'s in amino acids $i$ and $j$
whereas $\sigma _{ij}$ is determined pair-by-pair so that the minimum
in the potential is located at the experimentally established native distance.
The non-native contacts are repulsive below $r_{ij}$ of 4 {\AA}.

The implicit solvent is described by the Langevin noise and damping terms.
The amplitude of the noise is controlled by the temperature, $T$. All
simulations were done at $k_BT=0.3\: \epsilon$, where $k_B$ is the
Boltzmann constant. Newton's equations of motion are solved by the
fifth order predictor-corrector algorithm. The model is considered
in the overdamped limit so that the characteristic time scale, $\tau$,
is of order 1 ns as argued in refs. \cite{Veitshans,Peclet}.
Stretching is implemented by attaching an elastic spring
to two amino acids. The spring constant used has a value of
$0.12\: \epsilon/\mathrm{\AA}^2$ which is close to the elasticity of experimental
cantilevers. One of the springs is anchored and
the other spring is moving with a constant speed, $v_p$.
Choices in the value of the spring constant have been found to affect the
look of the force-displacements patterns and thus the location
of the transition state \cite{Ritchie,Seifert}, but not the values
of $F_{max}$ \cite{Hoang3,Hoang,JPCM}.

The dependence on $v_p$ is protein-dependent and it is
approximately logarithmic in $v_p$ as evidenced by Figure \ref{ekstra}
for several strong proteins. The logarithmic dependence
has been demonstrated experimentally, for instance, for
polyubiquitin \cite{C4,C2}.
$F_{max}\;=\;p\;ln(v/v_0)\;+\;q$.
The  approximate validity of this relationship is demonstrated
in Figure \ref{ekstra} for three proteins with big values of
$F_{max}$.
We observe that the larger the value of $F_{max}$,
the bigger probability that the dependence on $v_p$ is large.
When we make a fit to
$F_{max}\;=\;p\;ln(v/v_0)\;+\;q$
for 1vpf, 1c4p, and 1j8s, we get the parameter $p$ to be equal to
$0.39 \pm 0.11$, $0.17 \pm 0.03$,
and $0.04 \pm 0.02\: \epsilon/\mathrm{\AA}$
respectively (the values of $q$ are $7.42 \pm 0.63$, $5.85 \pm 0.16$,
and $4.96 \pm 0.08\: \epsilon/\mathrm{\AA}$ correspondingly).
However, some strong proteins may have $p$ to be as low as 0.04.

When making the survey, we have used $v_p$ of 0.005 $\mathrm{\AA}/\tau$
and stretching was accomplished by attaching the springs to the
terminal amino acids (there is an astronomical number of other choices
of the attachment points).

In order to estimate an effective experimental value of the
energy parameter $\epsilon$, we have correlated the theoretical
values of $F_{max}$ with those obtained experimentally.
The experimental data points used in ref. \cite{models} have
been augmented by entries pertaining to 1emb (117-182), 1emb (182-212)
\cite{newDietz} (where the numbers in brackets indicate the
amino acids that are pulled) and 1aoh, 1g1k, and 1amu \cite{Valbuena}.
The full list of the experimental entries is provided by Table 6.
Unlike the previous plots \cite{models} that cross correlate
the experimental and theoretical values of $F_{max}$, we now extrapolate
the theoretical forces to the values that should be measured at
the pulling speeds that are used experimentally.
We assume that the unit of speed, $v_0=1\; \mathrm{\AA}/\tau$, is
of order 1 {\AA}/ns and consider 10 speeds to make a
fit to the logarithmic relationship.
The values of parameters $p$ and $q$ for the proteins
studied experimentally are listed in Table 6.

The main panel of Figure \ref{exper} demonstrates the relationship
between the extrapolated theoretical and experimental values of
$F_{max}$. The best slope, indicated by the solid line, corresponds
to the slope of 0.0091. The inverse of this slope yields
110 pN as an effective equivalent of the theoretical force unit
of $\epsilon/\mathrm{\AA}$. The Pearson correlation coefficient, $R^2$ is
0.832, the rms percent error, $r_e$, is 1.02, and the Theil
$U$ coefficient (discussed in ref. \cite{models}) is 0.281.
The inset show a similar plot obtained when
the extrapolation to the experimental speeds is not done.
The resulting unit of the force would be equivalent to 110 pN
which differs form the previous estimate of 71 pN (shown by the
dotted line in the main panel)
because of the inclusion of
the newly measured proteins and implementation of the extrapolation
procedure.
The statistical measures of error here are $R^2=0.851$, $r_e=0.37$,
and $U=0.251$. These measures are better compared to the
case with the extrapolation because the extrapolation procedure
itself brings in additional uncertainties. Nevertheless,
implementing the procedure seems sounder physically.
The spread between these
various effective units of the force
suggests an error bar of order 30 pN
on the currently best value of 110 pN.
}

\section*{Acknowledgments}
{The idea of making surveys of proteins using Go-like models arose in a
very stimulating discussion with J. M. Fernandez in 2005. More recent
discussions and suggestions by M. Carrion-Vazquez, particularly
about the cysteine knots, are warmly appreciated.
}


\newpage
\clearpage
\section*{Tables}

TABLE 1: The predicted list of the strongest proteins.\\
\begin{centering}
\begin{small}
\begin{xtabular}{|l|l|l|l|l|l|l|l|} \hline
n & PDBid & N & $F_{max}\: [\epsilon/\mathrm{\AA}]$ & $L_{max} [\mathrm{\AA}]$& $\lambda$ & CATH & SCOP \\ \hline
 1 &\bf{ 1bmp }& 104 &\bf{~~ 10.2 }&  23.2 & 0.01 & 2.10.90.10 & g.17.1.2 \\
 2 &\bf{ 1qty }& 95 &\bf{~~ 8.9 }&  72.1 & 0.11 & 2.10.90.10 & b.1.1.4 \\
 3 &\bf{ 2bhk }& 119 &\bf{~~ 7.3 }&  26.5 & 0.67 &   &   \\
 4 &\bf{ 1lxi }& 104 &\bf{~~ 7.3 }&  22.5 & 0.01 &   & g.17.1.2 \\
 5 &\bf{ 1cz8 }& 107 &\bf{~~ 6.4 }&  76.5 & 0.13 & 2.10.90.10 & b.1.1.1 \\
 6 &\bf{ 2gh0 }& 219 &\bf{~~ 5.8 }&  25.9 & 0.06 &   &   \\
 7 &\bf{ 1wq9 }& 100 &\bf{~~ 5.5 }&  72.0 & 0.10 & 2.10.90.10 & g.17.1.1 \\
 8 &\bf{ 1flt }& 107 &\bf{~~ 5.5 }&  75.6 & 0.12 & 2.10.90.10 & b.1.1.4 \\
 9 &\bf{ 1fzv }& 117 &\bf{~~ 5.4 }&  90.4 & 0.12 & 2.10.90.10 & g.17.1.1 \\
 10 &\bf{ 2gyz }& 100 &\bf{~~ 5.4 }&  14.4 & 0.01 &   &   \\
 11 &\bf{ 1rew }& 103 &\bf{~~ 5.3 }&  21.7 & 0.01 & 2.10.90.10 & g.7.1.3 \\
 12 &\bf{ 1m4u }& 139 &\bf{~~ 5.3 }&  52.1 & 0.07 & 2.10.90.10 & g.17.1.2 \\
 13 &\bf{ 1vpf }& 94 &\bf{~~ 5.3 }&  68.1 & 0.11 & 2.10.90.10 & g.17.1.1 \\ \hline \hline
 14 &\bf{ 1c4p }& 137 &\bf{~~ 5.1 }& 106.0 & 0.12 & 3.10.20.180 & d.15.5.1 \\
 15 &\bf{ 1qqr }& 138 &\bf{~~ 5.0 }& 110.3 & 0.12 & 3.10.20.180 & d.15.5.1 \\
 16 &\bf{ 3bmp }& 114 &\bf{~~ 5.0 }&  33.0 & 0.03 & 2.10.90.10 & g.17.1.2 \\
 17 &\bf{ 1j8s }& 193 &\bf{~~ 4.9 }&  77.9 & 0.03 & 2.60.40.1370 & b.2.3.3 \\
 18 &\bf{ 1wq8 }& 96 &\bf{~~ 4.9 }&  82.6 & 0.11 & 2.10.90.10 & g.17.1.1 \\
 19 &\bf{ 1j8r }& 193 &\bf{~~ 4.8 }&  77.7 & 0.03 & 2.60.40.1370 & b.2.3.3 \\
 20 &\bf{ 1f3y }& 165 &\bf{~~ 4.8 }& 284.7 & 0.43 & 3.90.79.10 & d.113.1.1 \\
 21 &\bf{ 2vpf }& 109 &\bf{~~ 4.7 }&  79.3 & 0.11 & 2.10.90.10 & g.17.1.1 \\
 22 &\bf{ 2h64 }& 105 &\bf{~~ 4.6 }&  29.4 & 0.03 &   & g.7.1.3 \\
 23 &\bf{ 1kdm }& 177 &\bf{~~ 4.6 }& 309.4 & 0.45 & 2.60.120.200 & b.29.1.4 \\
 24 &\bf{ 1q56 }& 195 &\bf{~~ 4.5 }& 473.2 & 0.62 & 2.60.120.200 & b.29.1.4 \\
 25 &\bf{ 1rv6 }& 94 &\bf{~~ 4.5 }&  67.7 & 0.11 & 2.10.90.10 & b.1.1.4 \\
 26 &\bf{ 1waq }& 104 &\bf{~~ 4.5 }&  20.1 & 0.01 &   &   \\
 27 &\bf{ 1reu }& 103 &\bf{~~ 4.5 }&  20.4 & 0.01 & 2.10.90.10 & g.17.1.2 \\
 28 &\bf{ 1tgj }& 112 &\bf{~~ 4.4 }&  45.9 & 0.07 & 2.10.90.10 & g.17.1.2 \\
 29 &\bf{ 2pbt }& 133 &\bf{~~ 4.4 }& 219.9 & 0.39 &   &   \\
 30 &\bf{ 2h62 }& 104 &\bf{~~ 4.4 }&  24.3 & 0.02 &   & g.7.1.3 \\
 31 &\bf{ 1tgk }& 112 &\bf{~~ 4.4 }&  44.6 & 0.07 & 2.10.90.10 & g.17.1.2 \\
 32 &\bf{ 2fzl }& 197 &\bf{~~ 4.4 }&  49.7 & 0.02 &   & c.37.1.19 \\
 33 &\bf{ 1qu0 }& 181 &\bf{~~ 4.3 }& 156.9 & 0.22 & 2.60.120.200 & b.29.1.4 \\
 34 &\bf{ 1f5f }& 172 &\bf{~~ 4.3 }& 186.2 & 0.28 & 2.60.120.200 & b.29.1.4 \\
 35 &\bf{ 1dzk }& 148 &\bf{~~ 4.3 }& 110.3 & 0.16 & 2.40.128.20 & b.60.1.1 \\
 36 &\bf{ 1aoh }& 147 &\bf{~~ 4.3 }&  77.1 & 0.01 & 2.60.40.680 & b.2.2.2 \\
 37 &\bf{ 1vsc }& 196 &\bf{~~ 4.3 }& 238.3 & 0.24 & 2.60.40.10 & b.1.1.3 \\
 38 &\bf{ 2c7w }& 96 &\bf{~~ 4.2 }& 184.2 & 0.45 & 2.10.90.10 &   \\
 39 &\bf{ 2gyr }& 97 &\bf{~~ 4.2 }&  27.1 & 0.05 & 2.10.90.10 &   \\
 40 &\bf{ 1dzj }& 148 &\bf{~~ 4.2 }& 111.0 & 0.16 & 2.40.128.20 & b.60.1.1 \\
 41 &\bf{ 2sak }& 121 &\bf{~~ 4.2 }&  76.0 & 0.10 & 3.10.20.130 & d.15.5.1 \\
 42 &\bf{ 2bzm }& 129 &\bf{~~ 4.2 }& 124.3 & 0.24 &   &   \\
 43 &\bf{ 2pq1 }& 134 &\bf{~~ 4.1 }& 222.6 & 0.39 &   &   \\
 44 &\bf{ 1nwv }& 129 &\bf{~~ 4.1 }& 129.8 & 0.13 & 2.10.70.10 & g.18.1.1 \\
 45 &\bf{ 1e5g }& 120 &\bf{~~ 4.1 }& 133.1 & 0.17 & 2.10.70.10 & g.18.1.1 \\
 46 &\bf{ 2ick }& 220 &\bf{~~ 4.1 }& 462.5 & 0.54 &   &   \\
 47 &\bf{ 1gvl }& 223 &\bf{~~ 4.1 }& 114.9 & 0.09 & 2.40.10.10 & b.47.1.2 \\
 48 &\bf{ 1tgs }& 225 &\bf{~~ 4.1 }& 122.3 & 0.10 & 2.40.10.10 & b.47.1.2 \\
 49 &\bf{ 1u20 }& 196 &\bf{~~ 4.0 }& 408.5 & 0.53 &   & d.113.1.1 \\
 50 &\bf{ 1cui }& 197 &\bf{~~ 4.0 }& 422.8 & 0.55 & 3.40.50.1820 & c.69.1.30 \\
 51 &\bf{ 1ffd }& 197 &\bf{~~ 4.0 }& 423.0 & 0.55 & 3.40.50.1820 & c.69.1.30 \\
 52 &\bf{ 1kdk }& 177 &\bf{~~ 4.0 }& 357.2 & 0.53 & 2.60.120.200 & b.29.1.4 \\
 53 &\bf{ 2icj }& 219 &\bf{~~ 4.0 }& 455.9 & 0.53 &   &   \\
 54 &\bf{ 3dd5 }& 194 &\bf{~~ 4.0 }& 403.3 & 0.53 &   &   \\
 55 &\bf{ 1cug }& 197 &\bf{~~ 4.0 }& 422.6 & 0.55 & 3.40.50.1820 & c.69.1.30 \\
 56 &\bf{ 1b0o }& 161 &\bf{~~ 4.0 }& 237.3 & 0.36 & 2.40.128.20 & b.60.1.1 \\
 57 &\bf{ 1xza }& 197 &\bf{~~ 4.0 }& 422.9 & 0.55 & 3.40.50.1820 & c.69.1.30 \\
 58 &\bf{ 1vcd }& 126 &\bf{~~ 4.0 }& 199.7 & 0.37 &   & d.113.1.1 \\
 59 &\bf{ 1cuw }& 197 &\bf{~~ 4.0 }& 422.9 & 0.55 & 3.40.50.1820 & c.69.1.30 \\
 60 &\bf{ 1xzi }& 197 &\bf{~~ 4.0 }& 422.9 & 0.55 & 3.40.50.1820 & c.69.1.30 \\
 61 &\bf{ 1cus }& 197 &\bf{~~ 4.0 }& 423.3 & 0.55 & 3.40.50.1820 & c.69.1.30 \\
 62 &\bf{ 1cuf }& 197 &\bf{~~ 4.0 }& 423.1 & 0.55 & 3.40.50.1820 & c.69.1.30 \\
 63 &\bf{ 2a7h }& 223 &\bf{~~ 4.0 }& 114.7 & 0.10 & 2.40.10.10 & b.47.1.2 \\
 64 &\bf{ 1cq3 }& 224 &\bf{~~ 4.0 }& 128.0 & 0.12 & 2.60.240.10 & b.27.1.1 \\
 65 &\bf{ 1ffc }& 197 &\bf{~~ 3.9 }& 421.6 & 0.55 & 3.40.50.1820 & c.69.1.30 \\
 66 &\bf{ 1vc9 }& 126 &\bf{~~ 3.9 }& 199.1 & 0.37 &   & d.113.1.1 \\
 67 &\bf{ 1cua }& 197 &\bf{~~ 3.9 }& 423.0 & 0.55 & 3.40.50.1820 & c.69.1.30 \\
 68 &\bf{ 1xzl }& 197 &\bf{~~ 3.9 }& 423.1 & 0.55 & 3.40.50.1820 & c.69.1.30 \\
 69 &\bf{ 2faw }& 250 &\bf{~~ 3.9 }& 250.8 & 0.25 &   &   \\
 70 &\bf{ 2vn5 }& 142 &\bf{~~ 3.9 }&  49.2 & 0.02 &   &   \\
 71 &\bf{ 1cux }& 197 &\bf{~~ 3.9 }& 421.5 & 0.55 & 3.40.50.1820 & c.69.1.30 \\
 72 &\bf{ 1cuh }& 197 &\bf{~~ 3.9 }& 421.6 & 0.55 & 3.40.50.1820 & c.69.1.30 \\
 73 &\bf{ 2dsd }& 195 &\bf{~~ 3.9 }& 429.7 & 0.56 &   &   \\
 74 &\bf{ 2f3c }& 221 &\bf{~~ 3.9 }& 113.5 & 0.10 & 2.40.10.10 & b.47.1.2 \\
 75 &\bf{ 1xzj }& 197 &\bf{~~ 3.9 }& 421.8 & 0.55 & 3.40.50.1820 & c.69.1.30 \\
 76 &\bf{ 1xzf }& 197 &\bf{~~ 3.9 }& 421.0 & 0.55 & 3.40.50.1820 & c.69.1.30 \\
 77 &\bf{ 2g7i }& 124 &\bf{~~ 3.9 }& 106.6 & 0.10 &   &   \\
 78 &\bf{ 1g1k }& 143 &\bf{~~ 3.9 }&  52.0 & 0.02 & 2.60.40.680 & b.2.2.2 \\
 79 &\bf{ 1cuc }& 197 &\bf{~~ 3.9 }& 421.3 & 0.55 & 3.40.50.1820 & c.69.1.30 \\
 80 &\bf{ 1xzk }& 197 &\bf{~~ 3.9 }& 422.5 & 0.55 & 3.40.50.1820 & c.69.1.30 \\
 81 &\bf{ 1i04 }& 159 &\bf{~~ 3.9 }& 231.7 & 0.34 & 2.40.128.20 & b.60.1.1 \\
3144 &\bf{ 1ubq }&  76 &\bf{~~ 2.2 }&  47.9 & 0.04 & 3.10.20.90 & d.15.1.1 \\
3580 &\bf{ 1tit }&  89 &\bf{~~ 2.1 }&  55.3 & 0.04 & 2.60.40.10 & b.1.1.4 \\

\end{xtabular}
\end{small}
\end{centering}

\vspace*{0.4cm}
Table 1.  $F_{max}$ is obtained within the $LJ3$ model at the pulling velocity
of 0.005 $\mathrm{\AA}/\tau$.
The first column indicates the ranking of a model protein, the second --
the PDB code, and the third -- the number of the amino acids that are present
in the structure used.
$L_{max}$ denotes the end-to-end distance at which
the maximum force arises. $\lambda$ is the corresponding dimensionless location
defined as $\lambda = (L_{max}-L_n)/(L_f-L_n)$, where $L_n$ is the native end-to-end distance
and $L_f$ corresponds to full extension. The last two columns give the leading
CATH and SCOP codes.
The survey is performed based strictly on the PDB-assigned structure
codes. It may happen that the structure of a protein has been determined
several times and then each of these determinations leads to its own
value of $F_{max}$.
In this case, one may derive the best estimate either by picking
the best resolved structure or by making (weighted) averages over
all related structures.

\newpage
\noindent
TABLE 2. Gene Ontology terms for the top 190 proteins.

\vspace*{0.5cm}

\begin{xtabular}{|c| c| c |c| c | } \hline
Domain                  & GO identifier      & Term name                            & No. of structures  & Example \\ \hline
\bf{Molecular function} & GO:0016787         & hydrolase activity                   & 90                 & 1f3y \\\hline
                        & GO:0003824         & catalytic activity                   & 70                 & 1gvl \\\hline
                        & GO:0004252         & serine-type endopeptidase activity   & 39                 & 1c4p \\\hline
                        & GO:0008083         & growth factor activity               & 25                 & 1bmp \\ \hline
\bf{Biological process} & GO:0006508         & proteolytic activity                 & 34                 & 2a7h \\\hline
                        & GO:0007586         & digestion                            & 32                 & 1bra \\ \hline
\bf{Cellular component} & GO:0005576         & extracellular region                 & 122                & 1vpf, 1aoh \\\hline
                        & GO:0005515         & protein binding                      & 70                 & 1bmp \\

\hline
\end{xtabular}

\vspace*{2cm}

\noindent
TABLE 3: CATH classes (C), architectures (A), and topologies (T) contributing to the top strength proteins. The percentages indicated in the
column denode by "Strong" are relative the top 190 proteins listed in
Table 1. X corresponds to proteins not listed in CATH.\\
\vspace*{0.5cm}
\begin{centering}
\begin{small}
\begin{xtabular}{|l l l| l l l l l l| l |} \hline
{\large \bf{C}}      & {\large{\bf A}}&{\large \bf{T}}& {\large Strong} &                 &                & {\large All}     &              &           & {\large Root name }  \\ \hline
{\large 2.}          &                &               & {\large 57.3\%} &                 &                & {\large 26.4\%}  &              &           & {\large Mainly $\beta$ } \\
                     &{\bf 2.10}      &               &                 & {\bf 17.3\%}    &                &                  & 2.0\%        &           & {\bf Ribbon}    \\
                     &                & 2.10.90       &                 &                 & 12.1\%         &                  &              & 0.3\%     & Cystine Knot Cytokines, subunit B\\ 
                     &                & 2.10.70       &                 &                 & 5.2\%          &                  &              & 0.1\%     & Complement Module, domain 1  \\ 
                     &{\bf 2.40}      &               &                 & {\bf 25.7\%}    &                &                  & 8.9\%        &           & {\bf $\beta$ Barrel}                    \\
                     &                & 2.40.10       &                 &                 & 21.5\%         &                  &              & 2.9\%     & Thrombin,subunit H                     \\
                     &{\bf 2.60}      &               &                 & {\bf 14.2\%}    &                &                  & 10.6\%       &           & {\bf Sandwich}                          \\
                     &                & 2.60.40       &                 &                 & 3\%            &                  &              & 7\%       & Immunoglobulin-like         \\ \hline
{\large 3.}          &                &               & {\large 26.8\%} &                 &                & {\large 25.8\%}  &              &           & {\large $\alpha-\beta$}                   \\
                     &{\bf 3.10}      &               &                 & {\bf 8.4\%}     &                &                  & 5.2\%        &           & {\bf Roll}                 \\
                     &                & 3.10.20       &                 &                 & 2.6\%          &                  &              & 1.3\%     & Ubiquitin-like (UB roll)         \\
                     &                & 3.10.130      &                 &                 & 5.7\%          &                  &              & 1.0\%     & P-30 Protein                      \\ 
                     &{\bf 3.40}      &               &                 & {\bf 17.9\%}    &                &                  & 9.4\%        &           & {\bf 3-Layer (aba) Sandwich}          \\
                     &                & 3.40.50       &                 &                 & 17.9\%         &                  &              & 5.6\%     & Rossmann fold                   \\\hline
{\large X}           &                &               & {\large 15.7\%} &                 &                & {\large 26.6 \%} &              &           &           \\ \hline
\end{xtabular}
\end{small}
\end{centering}

\newpage

\noindent
TABLE 4: SCOP classes (C) and folds (F) contributing to the top strength proteins. X corresponds to proteins not listed in SCOP.\\
\vspace*{0.5cm}
\begin{centering}
\begin{small}
\begin{xtabular}{|ll|llll|p{3.3cm}|p{4cm}|}
\hline
{\bf C}     & {\bf F}     & {\large Strong}  &              & {\large All}      &                   & {\large Root name}                                                      &{\large Description} \\ \hline
{\large b.}    &             & {\large 40.5\%}  &              & {\large 22.7\%}   &                   & {\large  ~$\beta$}                                                      &  \\
               & {\bf b.47}  &                  & {\bf 21.5}\% &                   & 2.7\%             & SMAD/FHA domain                                                         & sandwich; 11 strands in 2 sheets; greek-key\\\hline
{\large c.}    &             &{\large 17.9\%}   &              & {\large 9\%}      &                   & {\large ~$\alpha/\beta$}                                                & Mainly parallel $\beta-\mathrm{sheets}$ ($\beta-\alpha-\beta$ units) \\
               & {\bf c.69}  &                  & {\bf 15.7}\% &                   & 0.3\%             & Pyruvate kinase C-terminal domain-like                                  & 3 layers: a/b/a; mixed $\beta-\mathrm{sheet}$ of 5 strands, order 32145, strand 5 is antiparallel to the rest \\\hline
{\large d.  }  &             &{\large 11.05\%}  &              & {\large 18.9\%}   &                   & {\large ~$\alpha+\beta$}                                                & Mainly antiparallel $\beta-\mathrm{sheets}$ (segregated $\alpha$ and $\beta$ regions) \\
               & {\bf d.5}   &                  & {\bf 5.8}\%  &                   & 0.9\%             & RNase A-like                                                            & contains long curved $\beta-\mathrm{sheet}$ and 3 helices\\
               & {\bf d.113} &                  & {\bf 2.6}\%  &                   & 0.2\%             & DsrC, the $\gamma$ subunit of dissimilatory sulfite reductase           & $\beta(3)-\alpha(5)$; meander $\beta-\mathrm{sheet}$ packed against array of helices \\\hline
{\large g.  }  &             &{\large 13.7\% }  &              & {\large 4.9\% }   &                   & {\large ~Small proteins}                                                & Usually dominated by metal ligand, heme, and/or disulfide bridges \\
               & {\bf g.17}  &                  & {\bf 5.2}\%  &                   & 0.1\%             & Necrosis inducing protein 1, NIP1                                       & disulfide-rich fold; $\mathrm{all}-\beta$; duplication: contains two structural repeats \\
               & {\bf g.18}  &                  & {\bf 6.3}\%  &                   & 0.2\%             & Trefoil/Plexin domain-like                                              & disulfide-rich fold; common core is $\alpha+\beta$ with two conserved disulfides \\\hline
{\large X }    &             & {\large 16.3\%}  &              & {\large  27.4\%}  &                   &                                                                         &  \\ \hline
\end{xtabular}

\end{small}
\end{centering}

\vspace*{2cm}
TABLE 5: Members of the cysteine-knot cytokines superfamilly.
VEGF stands for vascular endothelial growth factor,
BMP for bone morphogenetic protein, and
TGF for transforming growth factor.
The star $*$ indicates  uncomplexed proteins.
\\
\vspace*{0.5cm}
\begin{centering}
\begin{small}
\xentrystretch{-0.60}
\begin{xtabular}{l l p{5cm} } \hline
family                              &  domain/complex                       &  PDB             \\ \hline
VEGF                                &                                       &  \\
                                    &  VEGF-A                               & $\mathrm{1vpf}^*$,$\mathrm{2vpf}^*$,1cz8,1bj1,1flt,1qty,1fpt, 1mjv,1mkg,1mkk \\
                                    &  VEGF-B                               & 2c7w  \\
                                    &  VEGF-F                               & 1wq9,1wq8,1rv6,1fzv \\
TGF                                 &                                       &  \\
                                    &  BMP7/ActRII                          & 1lx5,1lxi, 1m4u, 1bmp \\
                                    &  BMP2/IA                              & 1reu, 1rew, 2es7, $\mathrm{3bmp}^*$ \\
                                    &  BMP2 ternary ligand-receptor complex &  2h62, 2h64 \\
                                    &  human arthemine/GFRbeta3             & 1tgj, 1tgk \\
                                    &  human arthemine/GFRalpha3            & 2gh0, 2gyz \\
BMP                                 &  human growth and differentiation factor 5 &  1waq , 2bhk \\
\hline
\hline
\end{xtabular}
\end{small}
\end{centering}

\clearpage
TABLE 6: The experimental and theoretical data on stretching of proteins.\\
\begin{centering}
\begin{small}
\xentrystretch{-0.60}
\begin{xtabular}{|c| c| c| c| c| c| c| c| r| p{1.2cm}|} \hline
n &PDB  &  $F_{max}^{e}$ [pN] &  $v_p$ [nm/s] & $F_{max}^{t} [\epsilon/\mathrm{\AA}]$&$F_{max}^{te} [\epsilon/\mathrm{\AA}]$&$\mathrm{p}[\epsilon/\mathrm{\AA}]$&$\mathrm{q}[\epsilon/\mathrm{\AA}]$& Note&Ref. \\ \hline
1&1tit &   204 +/-   30  & 600 &   2.15          & 1.85 & 0.040 & 2.335       &  I27*8    &    \cite{R4,C6}   \\ 
2&1nct &   210 +/-   10  & 500  &   2.4 +/- 0.2  & 1.48 & 0.100 & 2.703       &  I54-I59  &  \cite{W1,W3}   \\
3&1g1c &   127 +/-   10  & 600  &   2.3 +/- 0.2  & 2.23 & 0.038 & 2.680       &   I5 titin &  \cite{L6}    \\
4&1b6i &   64 +/-    30 & 1000  &   1.2          & 0.74 & 0.084 & 1.710       & T4 lysozyme(21-141) & \cite{Y1}   \\ 
5&1aj3 &    68 +/-    20 & 3000 &  1.23          & 0.71 & 0.107 & 1.830       &   spectrin R16   & \cite{Lenne_2000}\\
6&1dqv &    60 +/-    15 & 600 &   1.5           & 0.58 & 0.147 & 2.349       & calcium binding C2A  &  \cite{C3}  \\  
7&1rsy &    60 +/-    15 & 600 &   1.7 +/- 0.2   & 1.48 & 0.040 & 1.962       & calcium binding C2A & \cite{C3} \\ 
8&1byn &    60 +/-    15 & 600 &   1.4           & 1.18 & 0.066 & 1.981       & calcium binding C2A  &   \cite{C3} \\ 
9&1cfc  & $<20$ & 600 &  0.55                    & 0.37 & 0.052 & 0.997       &  calmodulin   &  \cite{C3} \\
10&1bni &    70  +/-   15 & 300 &  1.4, 1.7       & 1.06 & 0.044 & 1.606      & barnase/i27  & \cite{B1}    \\  
11&1bnr &    70  +/-   15 & 300 &   1.05          & 0.71 & 0.053 & 0.053      & barnase/i27 &  \cite{B1}    \\  
12&1bny &    70  +/-   15 & 300 &   1.1, 1.3      & 0.65 & 0.046 & 0.046      & barnase/i27 &\cite{B1}   \\   
13&1hz6 &   152  +/-   10 & 700 &   3.5           & 2.79 & 0.064 & 3.542      & protein L &   \cite{B3}  \\   
14&1hz5 &   152  +/-   10 & 700 &   2.8           & 2.22 & 0.104 & 0.104      & protein L &  \cite{B3} \\    
15&2ptl &   152  +/-   10 & 700 &   2.2 +/- 0.2   & 1.88 & 0.045 & 0.045      & protein L   & \cite{B3} \\
16&1ubq &   230  +/-   34 & 1000 &   2.32         & 1.47 & 0.134 & 3.019      & ubiquitin  & \cite{C2}  \\ 
17&1ubq &    85  +/-   20 & 300 &   0.9           & 0.72 & 0.083 & 1.779      & ubiquitin(K48-C)*(2-7)  &\cite{C4,C2} \\ 
18&1emb &   350  +/-   30 & 3600 &  5.15 +/- 0.4  & 4.16 & 0.121 & 5.403      & GFP(3-132) &\cite{Dietz_2004}\\  
19&1emb &   407  +/-   45 & 12000 &  5.15 +/- 0.4 & 4.30 & 0.121 & 5.403      & GFP(3-132) &\cite{Dietz_2006}\\  
20&1emb &   346  +/-   46 & 2000 &  5.15 +/- 0.4  & 4.09 & 0.121 & 5.403      & GFP(3-132) &\cite{Dietz_2006}\\  
21&1emb &   117  +/-   19 & 3600 &  2.3, 4.3      & 1.91 & 0.050 & 2.427      & GFP(3-212) & \cite{Dietz_2006}\\
22&1emb &   127  +/-   23 & 3600 &  2.2 +/- 0.2   & 1.51 & 0.164 & 3.197      & GFP(132-212) & \cite{Dietz_2006}\\  
23& 1emb & 548 +/-  57 & 3600 & 3.5 +/- 0.1       & 2.89 & 0.142 & 4.347      & GFP(117-182) &  \cite{newDietz} \\
24& 1emb & 356 +/- 61& 3600 & 3.2 +/- 0.2         & 2.94 & 0.075 & 3.709      & GFP(182-212) &  \cite{newDietz} \\
25&1emb &   104  +/-   40 & 3600 &   2.3 +/- 0.2  & 1.26 & 0.236 & 3.683      & GFP(N-C) & \cite{Dietz_2004}\\
26&1fnf &    75  +/-   20 & 3000 &   1.6, 1.8     & 1.70 & 0.130 & 3.069      & Fniii-10  & \cite{L5,O2} \\ 
27&1ttf &    75  +/-   20 & 600 &   0.7, 1.2      & 0.99 & 0.006 & 1.071      & Fniii-10 &\cite{O4} \\ 
28&1ttg &    75  +/-   20 & 600 &   0.7, 1.0      & 0.17 & 0.099 & 1.365      & Fniii-10 &\cite{O4} \\ 
29&1fnh &   124  +/-   18 & 600 &   1.8           & 1.10 & 0.127 & 2.635      & Fniii-12 &\cite{O2} \\ 
30&1fnh &   89  +/-   18 & 600 &   1.4, 1.7       & 1.10 & 0.127 & 2.635      & Fniii-13  &\cite{O2}\\  
31&1oww &   220  +/-   31 & 600 &   2.1 +/- 0.2   & 2.01 & 0.024 & 2.300      &FNiii-1  & \cite{O2} \\ 
32&1ten &  135  +/-   40 & 500 &   1.7            & 1.53 & 0.026 & 1.857      & TNFNiii-3 & \cite{O1,O2} \\ 
33&1pga &  190  +/-   20 & 400 &   2.4, +/- 0.2   & 2.50 & 0.001 & 2.761      & protein G &\cite{Hongbin}\\ 
34&1gb1 &  190  +/-   20 & 400 &   1.65 +/- 0.2   & 1.69 & 0.045 & 2.237      & protein G &\cite{Hongbin} \\ 
35&1aoh & 480 +/- 14 & 400 & 4.3 +/- 0.2          & 3.69 & 0.119 & 0.119      & scaffoldin c7A & \cite{Valbuena} \\
36&1g1k & 425 +/- 9 & 400 & 3.9 +/- 0.01          & 3.22 & 0.028 & 4.106      & scaffoldin c1C & \cite{Valbuena} \\
37&1anu & 214 +/- 8 & 400 & 3.3 +/- 0.03          & 2.55 & 0.060 & 3.224      & scaffoldin c2A &  \cite{Valbuena} \\
38&1qjo &    15 +/-    10 & 600 &   1.2           & 1.25 & 0.029 & 1.601      &  eE2lip3(N-C)  & \cite{B2}   \\
\hline
\end{xtabular}
\end{small}
\end{centering}

\vspace*{1cm}
Table 6.  $F^e_{max}$ denotes the experimentally measured value of
$F_{max}$ as reported in the reference stated in the last column.
$v_p$ denotes the experimental pulling speed used. $F^t_{max}$
is the value of the maximal force obtained in our simulation
within the $LJ3$ model. They were performed at
$v_p=0.005 \mathrm{\AA}/\tau$.
$F^{te}_{max}$ corresponds to the theoretical estimate of $F_{max}$
when extrapolated to the experimental speeds. The extrapolation assumes
the approximate logarithmic dependence $F_{max}\;=\;p\; ln(v/v_o) \;+\;q$,
where $v_0$ is $1\: \mathrm{\AA}/\tau$.
10 speeds were used to determine the values of $p$ and $q$ in analogy
to the procedure illustrated in Figure \ref{ekstra}
The values of $p$ and $q$ are provided
in columns 7 and 8 of the Table respectively. The first column
indicates the corresponding symbol that is used in Figure \ref{exper}.

\clearpage
\section*{Figure Legends}

\begin{figure}[!ht]
\epsfxsize=5in
\centerline{\epsffile{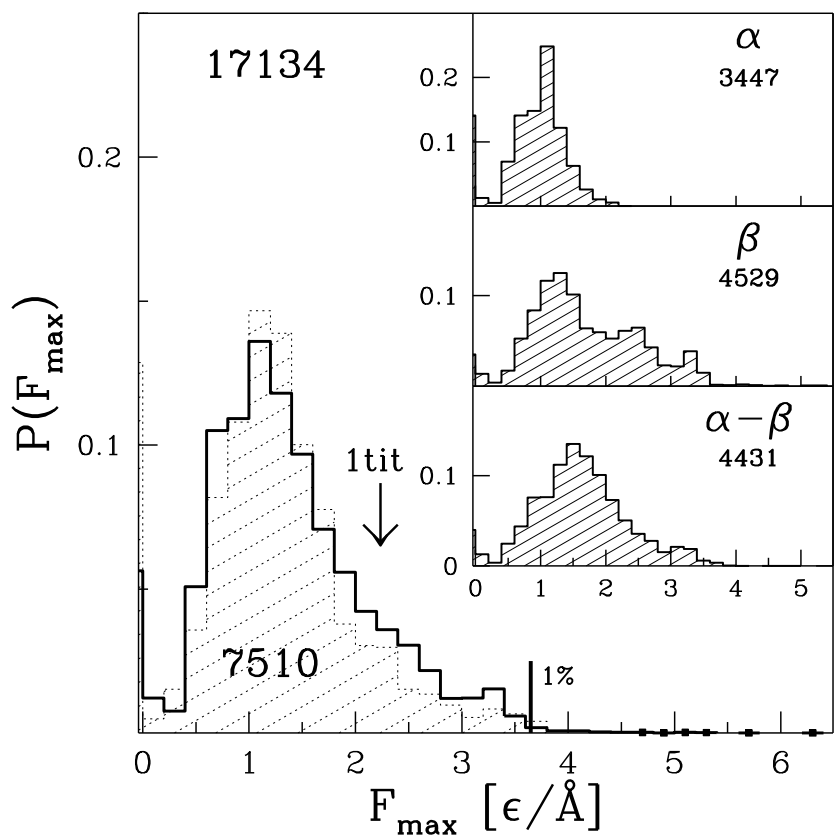}}
\vspace*{3cm}
\caption{{\bf Probability distribution of the maximal forces obtained
in the set of 17 134 model proteins (solid line).} The shaded
histogram corresponds to the 7510 proteins studied in ref. \cite{BJ}.
The insets show similar distributions for the CATH-based classes
indicated. The numbers underneath the class symbols give the size of
the set of the proteins considered. }
\label{allhist}
\end{figure}

\begin{figure}[!ht]
\epsfxsize=7in
\centerline{\epsffile{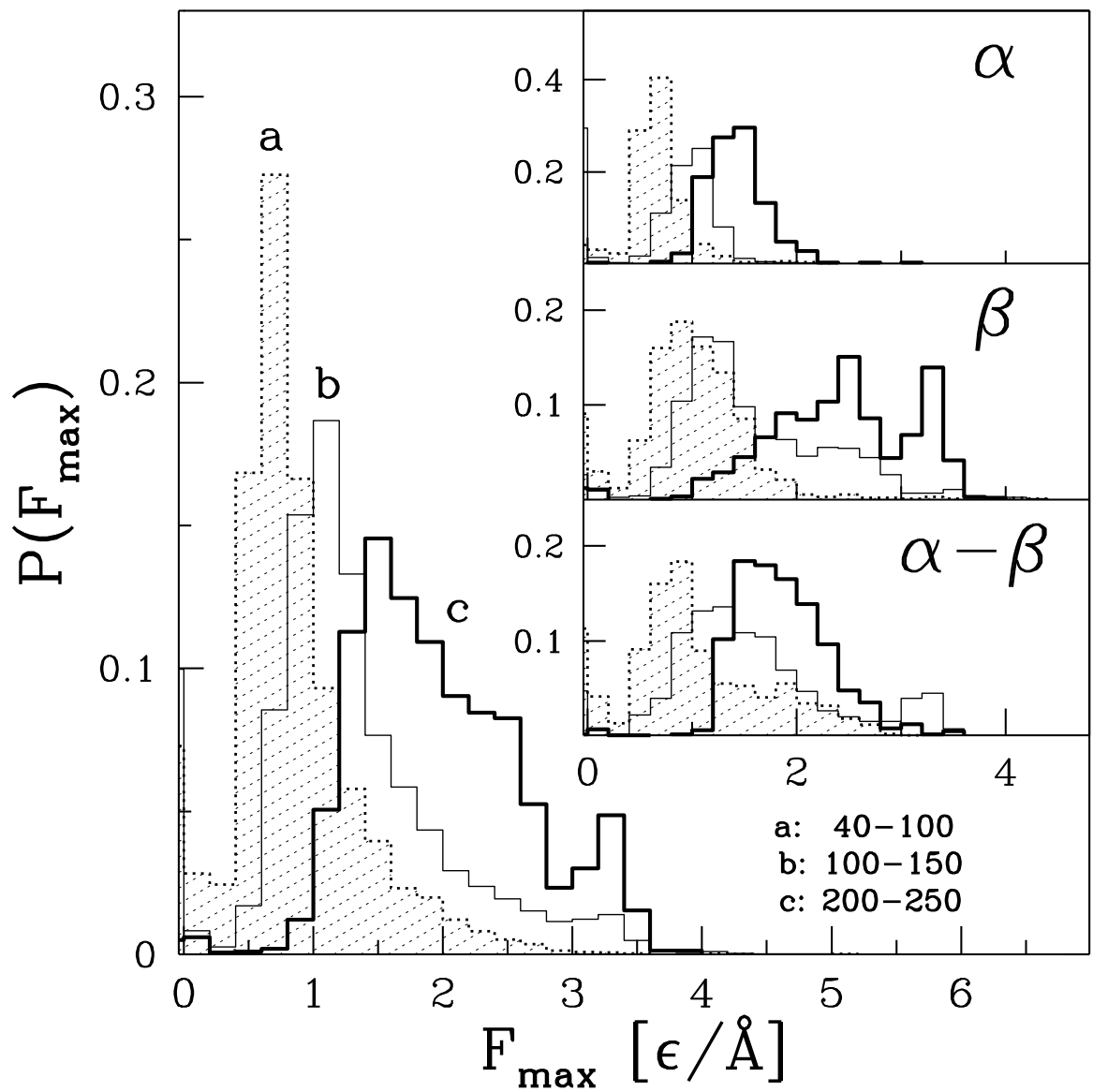}}
\vspace*{3cm}
\caption{{\bf Similar to Figure \ref{allhist} but for proteins belonging
to specific ranges of the sequential sizes, as indicated by
the symbols a, b, and c.}}
\label{sizedep}
\end{figure}

\begin{figure}[!ht]
\epsfxsize=7in
\centerline{\epsffile{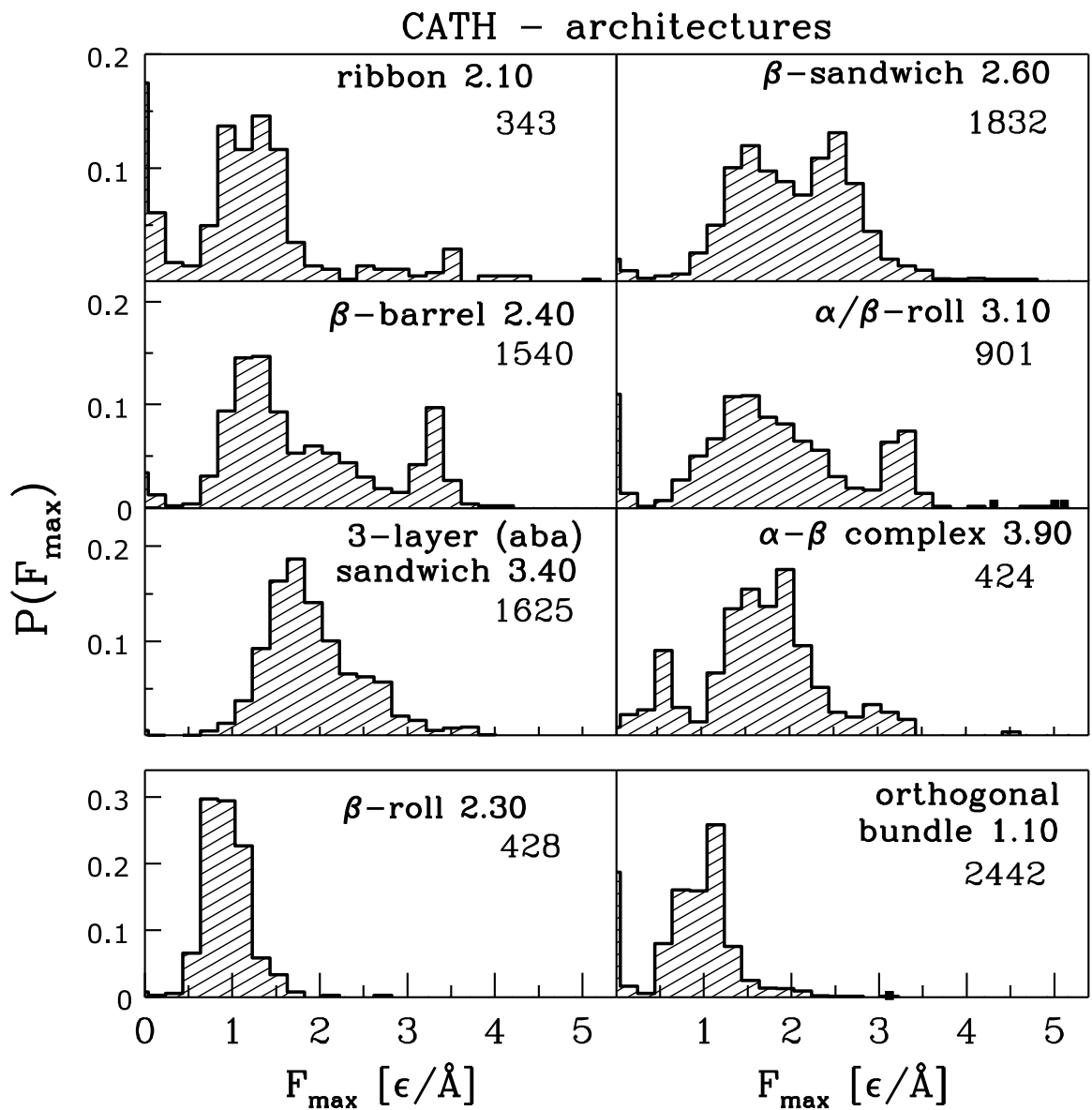}}
\vspace*{3cm}
\caption{{\bf The top six panels show probability distributions of $F_{max}$ for the
architectures that contribute to the pool of proteins with large forces.}
The architectures are indicated by their names and the accompanying
CATH numerical symbol. The numbers underneath the symbols of the
architecture inform about the number of cases contributing to the
distribution. The bottom two panels show examples of architectures that are predicted
to yield only small values of $F_{max}$.}
\label{catharch}
\end{figure}

\begin{figure}[!ht]
\epsfxsize=7in
\centerline{\epsffile{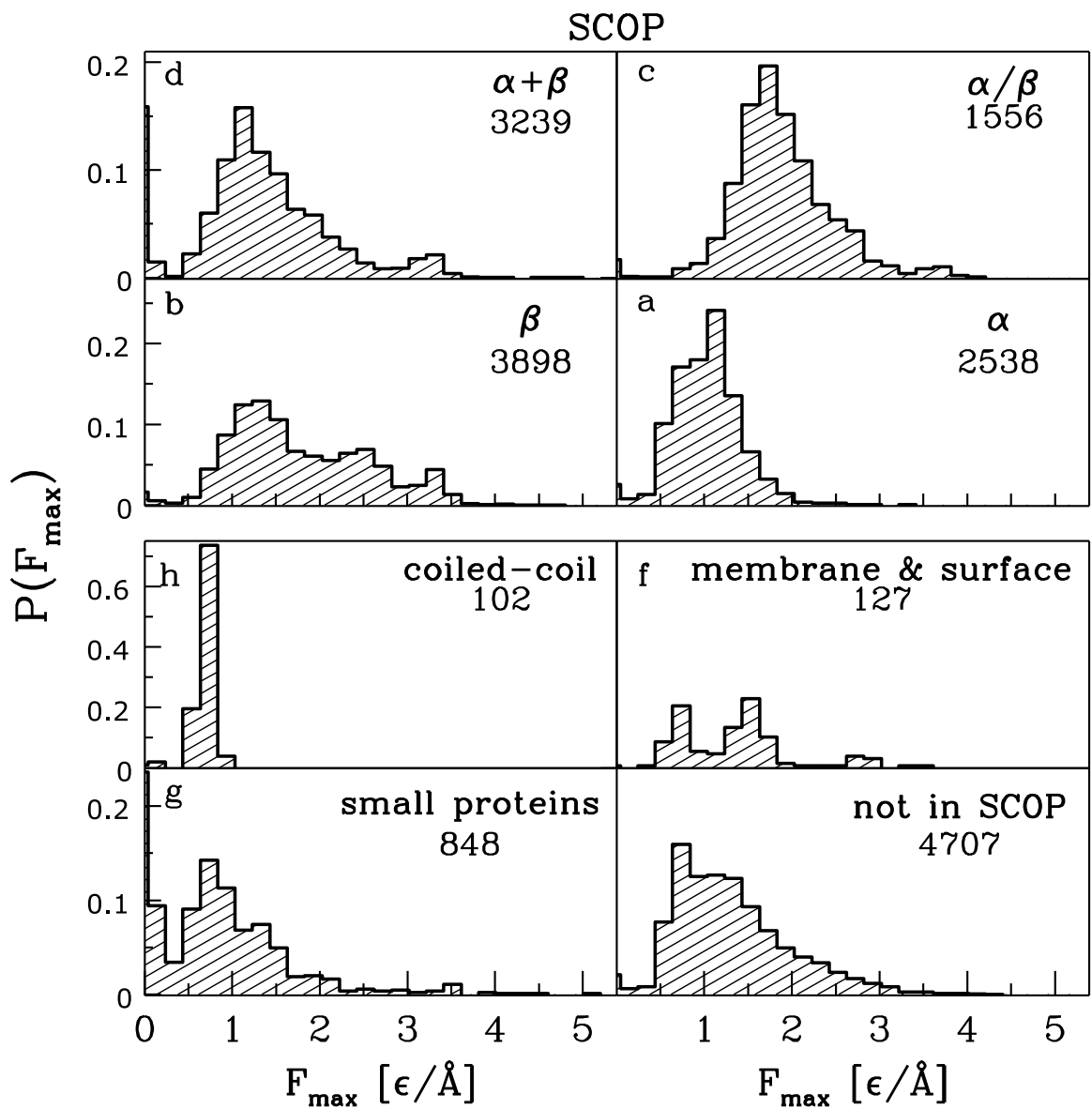}}
\vspace*{3cm}
\caption{{\bf Distributions of $F_{max}$ for the SCOP-based classes
for which there are more than 60 structures that could be used
in molecular dynamics studies.} The cases that are not shown are:
class e (27 structures), quasi-class i (5 structures),
and quasi-class j (52 structures). The bottom right
panel corresponds to structures which have no assigned
SCOP-based structure label. The numbers indicate the corresponding
numbers of structures studied.}
\label{scop}
\end{figure}

\begin{figure}[!ht]
\epsfxsize=7in
\centerline{\epsffile{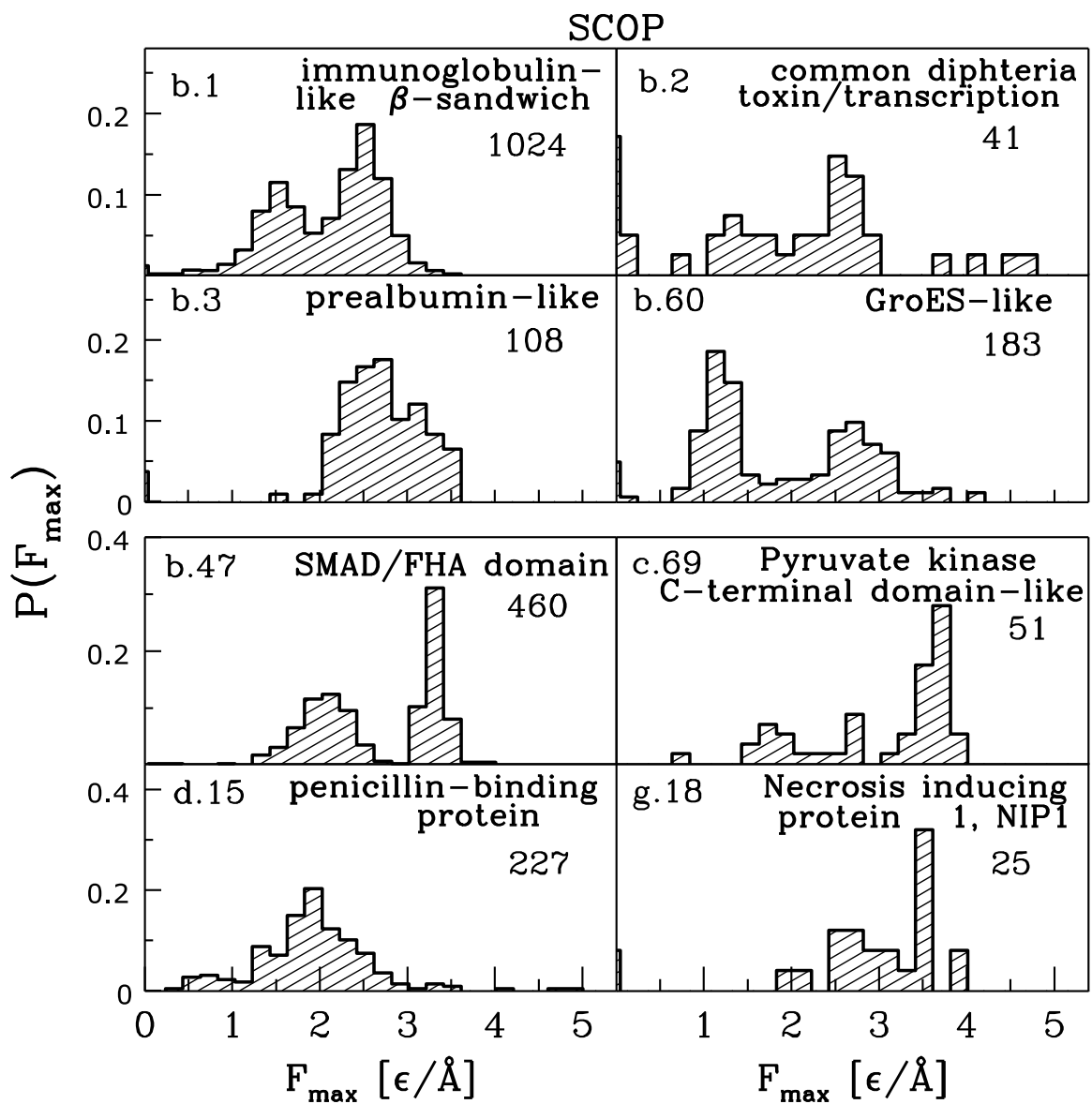}}
\vspace*{3cm}
\caption{{\bf Distributions of $F_{max}$  for eight folds
that may give rise to a large resistance to pulling.}}
\label{scop1}
\end{figure}

\begin{figure}[!ht]
\epsfxsize=7in
\centerline{\epsffile{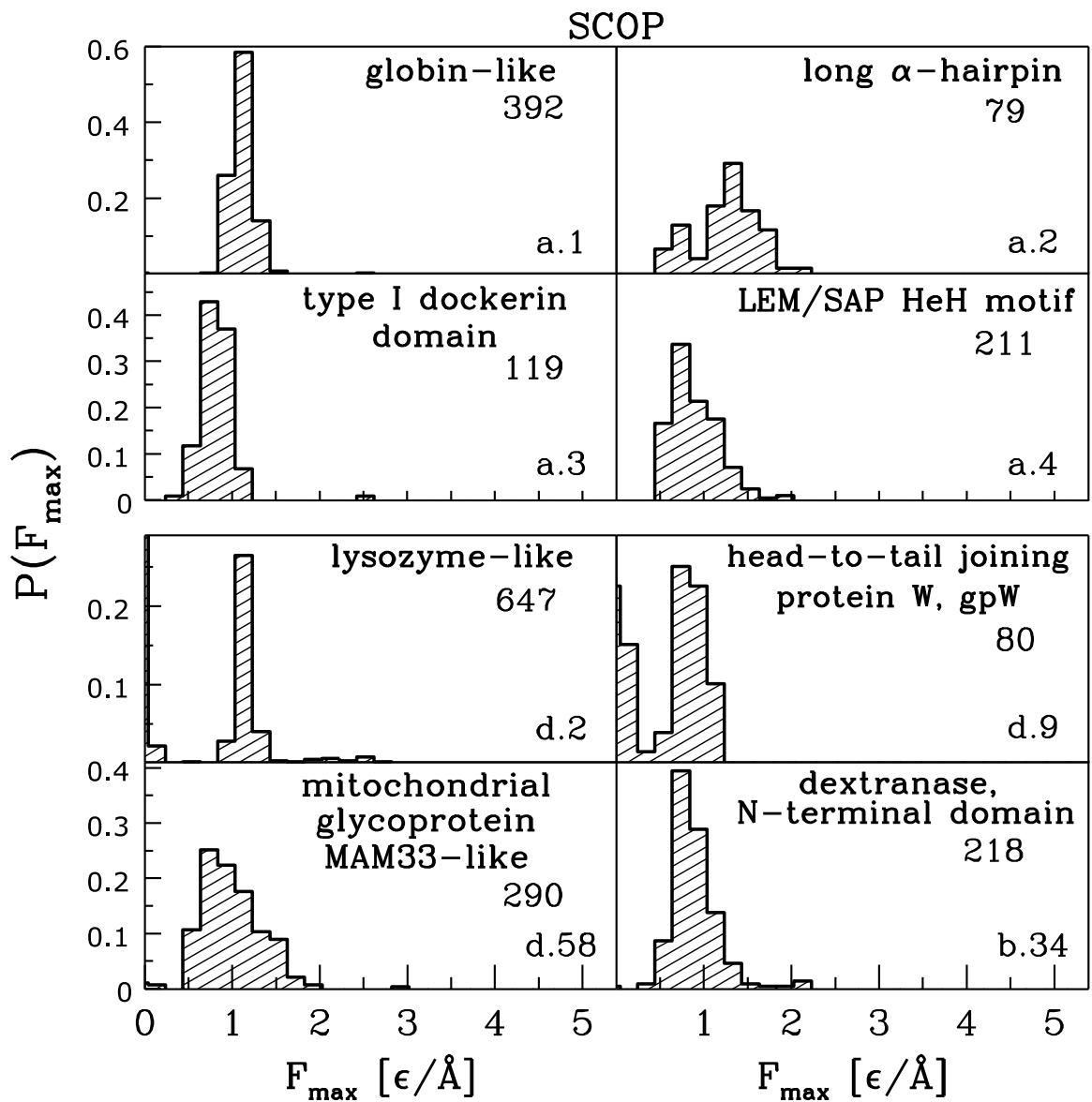}}
\vspace*{3cm}
\caption{{\bf Distribution of $F_{max}$ for eight folds that
are likely to yield a small resistance to pulling.}}
\label{scop2}
\end{figure}

\begin{figure}[!ht]
\epsfxsize=7in
\centerline{\epsffile{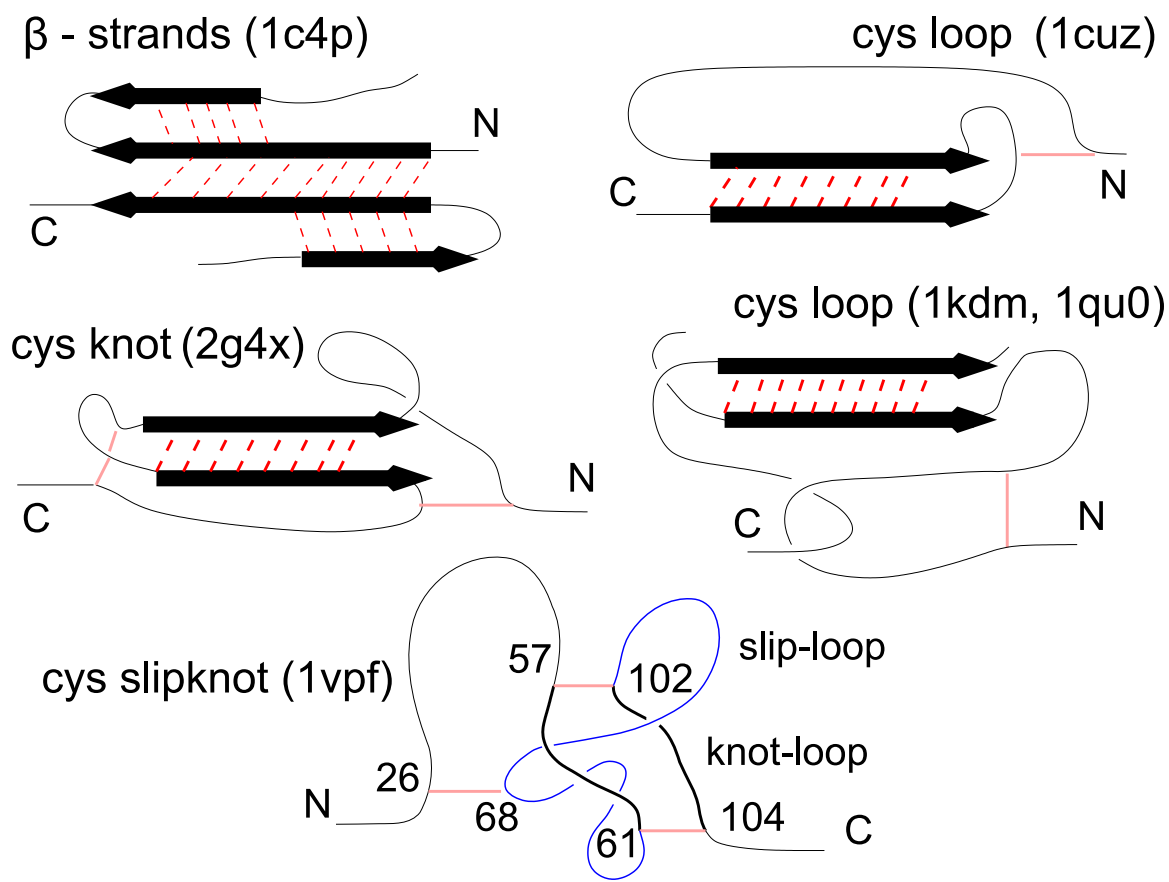}}
\vspace*{3cm}
\caption{{\bf Examples of force clamps found in the top strength proteins.} The relevant
disulphide bonds are shown in gray shade. The PDB codes of the examples of the
proteins that show the particular type of a clamp are indicated. In the case of the
CSK, the numbers indicate sequential locations of the amino acids
participating in a disulphide bridge in the 13-ranked 1vpf.}
\label{cysclamp}
\end{figure}

\begin{figure}[!ht]
\epsfxsize=7in
\centerline{\epsffile{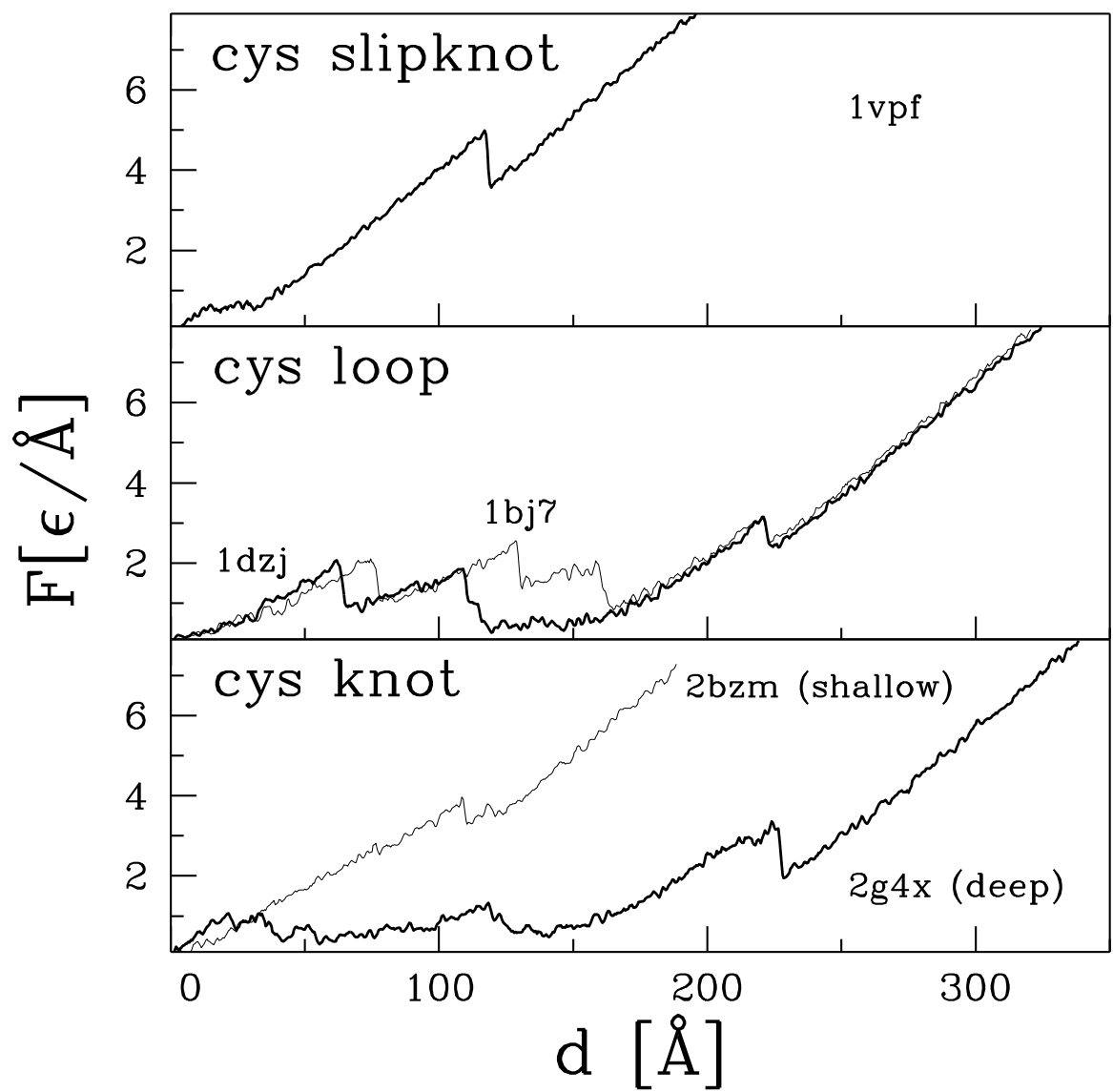}}
\vspace*{3cm}
\caption{{\bf Examples of the force patterns corresponding to proteins with the disulphide bonds.}}
\label{deep}
\end{figure}

\begin{figure}[!ht]
\epsfxsize=4in
\centerline{\epsffile{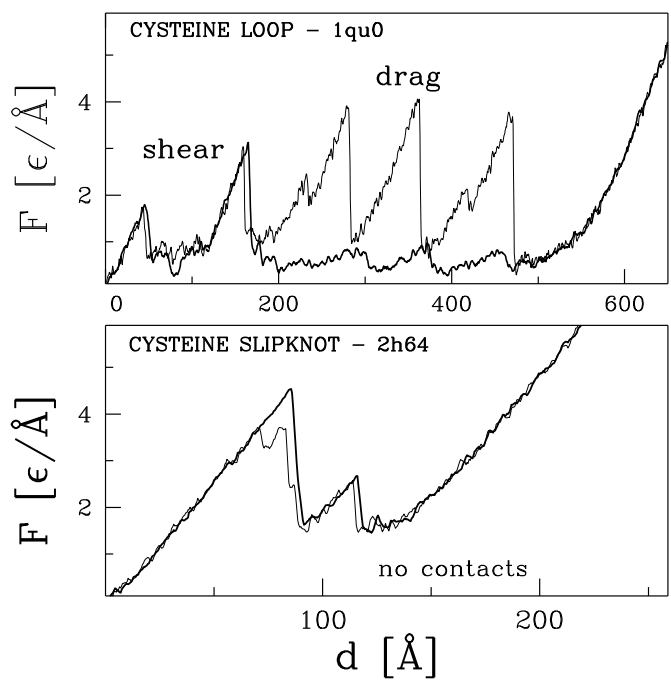}}
\caption{{\bf Top: Two trajectories arising in protein 1qu0.} Dragging occurs when the backbone
is pulled across the cysteine loop. Shearing occurs when the pull across the cystein loop
does not take place.
Bottom: The force-displacement pattern corresponding to the CSK  force
clamp in 2h64 (thick line). The thin line shows the corresponding pattern when one
removes the attractive contacts that are slipknot related.}
\label{pierscien1}
\end{figure}

\begin{figure}[!ht]
\epsfxsize=7in
\centerline{\epsffile{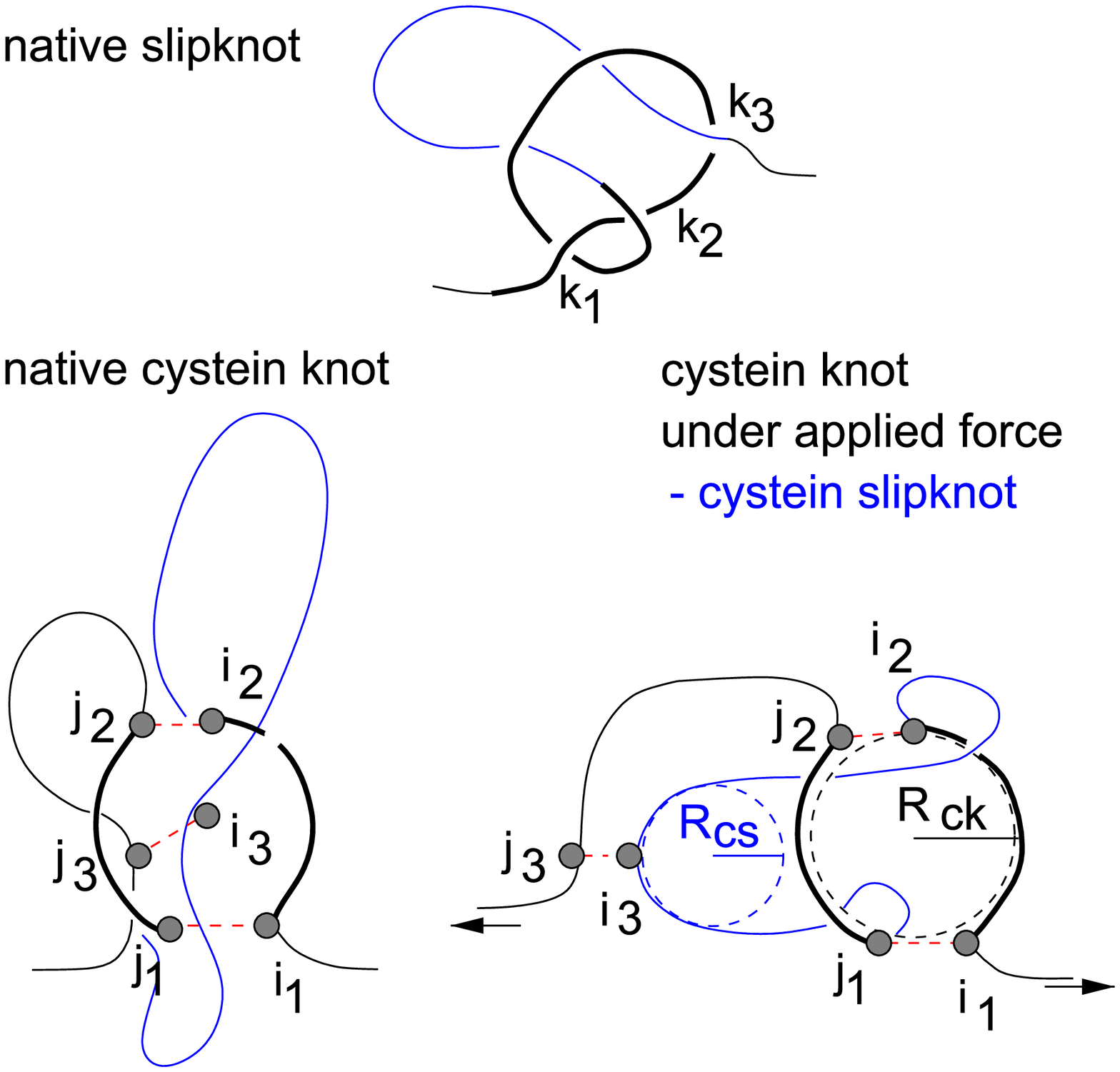}}
\vspace*{3cm}
\caption{{\bf Geometry of a slipknot and a cystein slipknot.} The top panel
corresponds to a genuine slipknot. The bottom left panel is a schematic representation
of the native geometry that yields the cystein slip-knot on stretching. The resulting
cystein slipknot motif is shown in the bottom right panel.}
\label{cskloop}
\end{figure}

\begin{figure}[!ht]
\epsfxsize=7in
\centerline{\epsffile{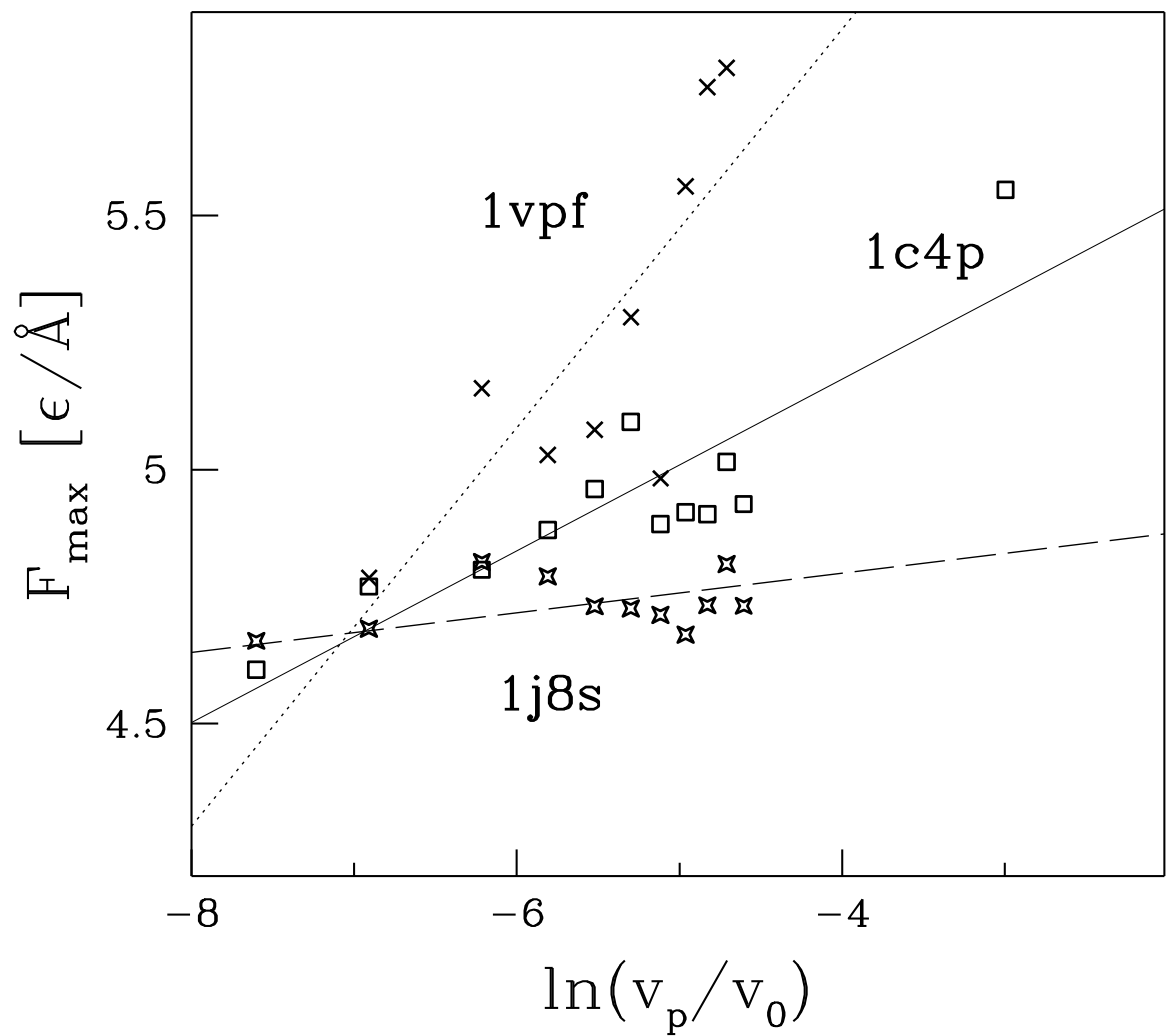}}
\vspace*{3cm}
\caption{{\bf Dependence of $F_{max}$ and the pulling velocity
for the proteins indicated.} $v_0$ corresponds to $1\: \mathrm{\AA}/\tau$
which is of order $10^8$ nm/s. The data for several top strength
proteins are shown.}
\label{ekstra}
\end{figure}

\begin{figure}[!ht]
\epsfxsize=6in
\centerline{\epsffile{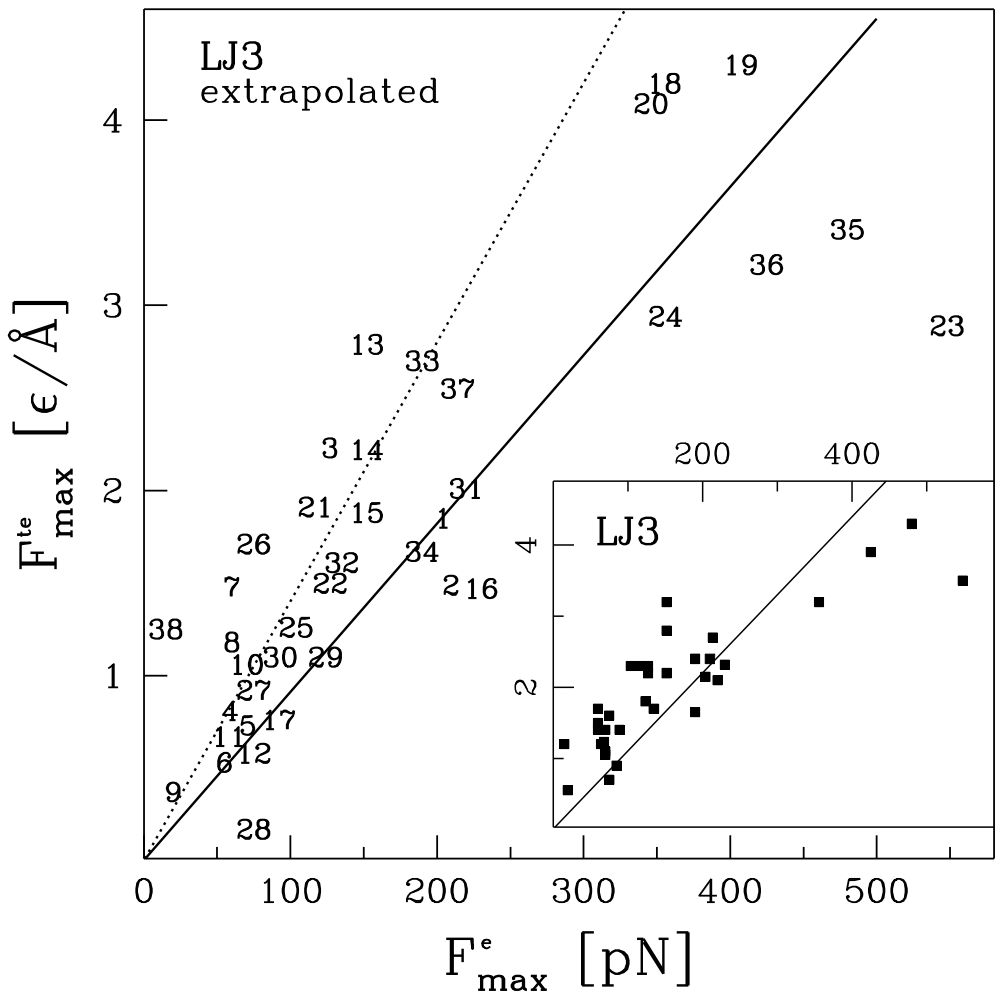}}
\caption{{\bf Theoretical $F^{te}_{max}$ extrapolated to the
pulling speeds used experimentally vs. the corresponding experimental
value, $F^e_{max}$.} The solid line indicates the best slope of
1/(110 pN). The dotted line corresponds to the previous result of
1/(71 pN) obtained in ref. \cite{models} where no
exptrapolation was made.
The inset shows a similar plot in which
the extrapolation is not implemented (denoted as $F^t_{max}$ in Table 6).
The list of the proteins used is provided by Table 6.
It comprises almost all cases considered in ref.
\cite{models} but it also includes the recent data points obtained
for the scaffoldin proteins \cite{Valbuena} and the GFP \cite{newDietz}.
The numerical symbols used in the Figure match the listing
number in Table 6.}

\label{exper}
\end{figure}

\newpage
\cleardoublepage
\begin{center}
{\bf SUPPLEMENTARY INFORMATION}
\end{center}

\section*{Tables}

TABLE 1S: The predicted list of the strongest proteins, ctd.\\
\begin{centering}
\begin{small}
\begin{xtabular}{|l|l|l|l|l|l|l|l|} \hline
n & PDBid & N & F$_{max}$ [$\epsilon /${\AA}] & $L_{max}$ [{\AA}]& $\lambda$ & CATH & SCOP \\ \hline
 82 &\bf{ 2duk }& 138 &\bf{~~ 3.8 }& 242.2 & 0.43 &   &   \\
 83 &\bf{ 1hqp }& 149 &\bf{~~ 3.8 }& 197.7 & 0.32 & 2.40.128.20 & b.60.1.1 \\
 84 &\bf{ 1cuu }& 197 &\bf{~~ 3.8 }& 421.4 & 0.55 & 3.40.50.1820 & c.69.1.30 \\
 85 &\bf{ 1afk }& 124 &\bf{~~ 3.8 }& 175.4 & 0.33 & 3.10.130.10 & d.5.1.1 \\
 86 &\bf{ 2o5w }& 147 &\bf{~~ 3.8 }& 214.8 & 0.33 &   &   \\
 87 &\bf{ 1xzc }& 197 &\bf{~~ 3.8 }& 422.9 & 0.55 & 3.40.50.1820 & c.69.1.30 \\
 88 &\bf{ 1qoz }& 206 &\bf{~~ 3.8 }& 238.7 & 0.29 & 3.40.50.1820 & c.69.1.30 \\
 89 &\bf{ 3pcf }& 200 &\bf{~~ 3.8 }& 251.2 & 0.31 & 2.60.130.10 & b.3.6.1 \\
 90 &\bf{ 1odi }& 234 &\bf{~~ 3.8 }& 458.8 & 0.51 & 3.40.50.1580 & c.56.2.1 \\
 91 &\bf{ 1y2x }& 142 &\bf{~~ 3.8 }&  38.1 & 0.01 & 2.60.270.20 & b.97.1.2 \\
 92 &\bf{ 1mbq }& 220 &\bf{~~ 3.8 }& 106.8 & 0.09 & 2.40.10.10 & b.47.1.2 \\
 93 &\bf{ 1bj7 }& 150 &\bf{~~ 3.8 }& 195.1 & 0.31 & 2.40.128.20 & b.60.1.1 \\
 94 &\bf{ 1odl }& 234 &\bf{~~ 3.8 }& 458.8 & 0.51 & 3.40.50.1580 & c.56.2.1 \\
 95 &\bf{ 1lx5 }& 104 &\bf{~~ 3.8 }&  21.7 & 0.01 & 2.10.90.10 & g.7.1.3 \\
 96 &\bf{ 1cuz }& 196 &\bf{~~ 3.8 }& 416.5 & 0.54 & 3.40.50.1820 & c.69.1.30 \\
 97 &\bf{ 3pch }& 200 &\bf{~~ 3.8 }& 252.3 & 0.31 & 2.60.130.10 & b.3.6.1 \\
 98 &\bf{ 1oxm }& 196 &\bf{~~ 3.8 }& 418.8 & 0.55 & 3.40.50.1820 & c.69.1.30 \\
 99 &\bf{ 1h2p }& 125 &\bf{~~ 3.8 }& 129.7 & 0.14 & 2.10.70.10 & g.18.1.1 \\
 100 &\bf{ 1gwy }& 175 &\bf{~~ 3.8 }& 143.2 & 0.18 & 2.60.270.20 & b.97.1.1 \\
 101 &\bf{ 1cud }& 197 &\bf{~~ 3.8 }& 420.1 & 0.55 & 3.40.50.1820 & c.69.1.30 \\
 102 &\bf{ 1vvd }& 118 &\bf{~~ 3.8 }& 112.1 & 0.14 & 2.10.70.10 & g.18.1.1 \\
 103 &\bf{ 1hfh }& 120 &\bf{~~ 3.8 }& 107.3 & 0.15 & 2.10.70.10 & g.18.1.1 \\
 104 &\bf{ 1vvc }& 118 &\bf{~~ 3.8 }& 112.9 & 0.14 & 2.10.70.10 & g.18.1.1 \\
 105 &\bf{ 1cuv }& 197 &\bf{~~ 3.8 }& 420.3 & 0.55 & 3.40.50.1820 & c.69.1.30 \\
 106 &\bf{ 1c77 }& 130 &\bf{~~ 3.8 }& 109.5 & 0.18 & 3.10.20.130 & d.15.5.1 \\
 107 &\bf{ 1xuk }& 223 &\bf{~~ 3.8 }& 115.0 & 0.10 & 2.40.10.10 & b.47.1.2 \\
 108 &\bf{ 1c2k }& 223 &\bf{~~ 3.8 }& 114.4 & 0.10 & 2.40.10.10 & b.47.1.2 \\
 109 &\bf{ 2stb }& 222 &\bf{~~ 3.8 }& 113.1 & 0.09 & 2.40.10.10 & b.47.1.2 \\
 110 &\bf{ 3tgi }& 223 &\bf{~~ 3.8 }& 110.5 & 0.10 & 2.40.10.10 & b.47.1.2 \\
 111 &\bf{ 3byr }& 88 &\bf{~~ 3.8 }& 196.8 & 0.57 &   &   \\
 112 &\bf{ 1a0j }& 223 &\bf{~~ 3.8 }& 111.6 & 0.10 & 2.40.10.10 & b.47.1.2 \\
 113 &\bf{ 2pcd }& 200 &\bf{~~ 3.7 }& 251.3 & 0.31 & 2.60.130.10 & b.3.6.1 \\
 114 &\bf{ 1vve }& 118 &\bf{~~ 3.7 }& 104.3 & 0.11 & 2.10.70.10 & g.18.1.1 \\
 115 &\bf{ 2pf6 }& 231 &\bf{~~ 3.7 }& 488.0 & 0.51 &   &   \\
 116 &\bf{ 3pcl }& 200 &\bf{~~ 3.7 }& 252.5 & 0.31 & 2.60.130.10 & b.3.6.1 \\
 117 &\bf{ 1afl }& 124 &\bf{~~ 3.7 }& 175.3 & 0.33 & 3.10.130.10 & d.5.1.1 \\
 118 &\bf{ 1bs9 }& 207 &\bf{~~ 3.7 }& 241.4 & 0.29 & 3.40.50.1820 & c.69.1.30 \\
 119 &\bf{ 1tpa }& 223 &\bf{~~ 3.7 }& 113.4 & 0.10 & 2.40.10.10 & b.47.1.2 \\
 120 &\bf{ 3rn3 }& 124 &\bf{~~ 3.7 }& 168.5 & 0.31 & 3.10.130.10 & d.5.1.1 \\
 121 &\bf{ 2grk }& 228 &\bf{~~ 3.7 }& 135.6 & 0.12 & 2.60.240.10 &   \\
 122 &\bf{ 1xzh }& 197 &\bf{~~ 3.7 }& 421.1 & 0.55 & 3.40.50.1820 & c.69.1.30 \\
 123 &\bf{ 1xui }& 223 &\bf{~~ 3.7 }& 113.7 & 0.10 & 2.40.10.10 & b.47.1.2 \\
 124 &\bf{ 1rpg }& 124 &\bf{~~ 3.7 }& 204.4 & 0.40 & 3.10.130.10 & d.5.1.1 \\
 125 &\bf{ 1xuj }& 223 &\bf{~~ 3.7 }& 115.3 & 0.10 & 2.40.10.10 & b.47.1.2 \\
 126 &\bf{ 1bra }& 223 &\bf{~~ 3.7 }& 112.1 & 0.10 & 2.40.10.10 & b.47.1.2 \\
 127 &\bf{ 1rtb }& 124 &\bf{~~ 3.7 }& 202.9 & 0.39 & 3.10.130.10 & d.5.1.1 \\
 128 &\bf{ 1c1o }& 223 &\bf{~~ 3.7 }& 113.8 & 0.10 & 2.40.10.10 & b.47.1.2 \\
 129 &\bf{ 1gkg }& 136 &\bf{~~ 3.7 }& 142.9 & 0.21 & 2.10.70.10 & g.18.1.1 \\
 130 &\bf{ 1c5v }& 223 &\bf{~~ 3.7 }& 115.0 & 0.10 & 2.40.10.10 & b.47.1.2 \\
 131 &\bf{ 1tnk }& 223 &\bf{~~ 3.7 }& 113.5 & 0.10 & 2.40.10.10 & b.47.1.2 \\
 132 &\bf{ 1tzh }& 94 &\bf{~~ 3.7 }&  67.4 & 0.10 & 2.10.90.10 & b.1.1.1 \\
 133 &\bf{ 1ckl }& 126 &\bf{~~ 3.7 }& 127.3 & 0.16 & 2.10.70.10 & g.18.1.1 \\
 134 &\bf{ 2fwu }& 157 &\bf{~~ 3.7 }&  88.2 & 0.10 &   & b.1.27.1 \\
 135 &\bf{ 1aqp }& 124 &\bf{~~ 3.7 }& 204.7 & 0.40 & 3.10.130.10 & d.5.1.1 \\
 136 &\bf{ 2g4x }& 124 &\bf{~~ 3.7 }& 205.1 & 0.40 & 3.10.130.10 & d.5.1.1 \\
 137 &\bf{ 2sta }& 222 &\bf{~~ 3.7 }& 114.3 & 0.10 & 2.40.10.10 & b.47.1.2 \\
 138 &\bf{ 1h03 }& 125 &\bf{~~ 3.7 }& 128.9 & 0.14 & 2.10.70.10 & g.18.1.1 \\
 139 &\bf{ 1mtv }& 223 &\bf{~~ 3.7 }& 112.2 & 0.10 & 2.40.10.10 & b.47.1.2 \\
 140 &\bf{ 1co7 }& 223 &\bf{~~ 3.7 }& 110.7 & 0.10 & 2.40.10.10 & b.47.1.2 \\
 141 &\bf{ 2o1c }& 147 &\bf{~~ 3.7 }& 215.3 & 0.33 &   &   \\
 142 &\bf{ 1anc }& 223 &\bf{~~ 3.7 }& 110.7 & 0.10 & 2.40.10.10 & b.47.1.2 \\
 143 &\bf{ 1utl }& 222 &\bf{~~ 3.7 }& 113.1 & 0.10 & 2.40.10.10 & b.47.1.2 \\
 144 &\bf{ 1btp }& 223 &\bf{~~ 3.7 }& 113.2 & 0.10 & 2.40.10.10 & b.47.1.2 \\
 145 &\bf{ 1xuh }& 223 &\bf{~~ 3.7 }& 115.4 & 0.10 & 2.40.10.10 & b.47.1.2 \\
 146 &\bf{ 1o72 }& 175 &\bf{~~ 3.7 }& 142.9 & 0.18 & 2.60.270.20 & b.97.1.1 \\
 147 &\bf{ 2ofc }& 141 &\bf{~~ 3.7 }&  37.7 & 0.01 & 2.60.270.20 &   \\
 148 &\bf{ 6rsa }& 124 &\bf{~~ 3.7 }& 203.7 & 0.39 & 3.10.130.10 & d.5.1.1 \\
 149 &\bf{ 2ofe }& 141 &\bf{~~ 3.7 }&  37.8 & 0.01 & 2.60.270.20 &   \\
 150 &\bf{ 2ofd }& 141 &\bf{~~ 3.7 }&  37.9 & 0.01 & 2.60.270.20 &   \\
 151 &\bf{ 2dsb }& 206 &\bf{~~ 3.7 }& 470.0 & 0.58 &   &   \\
 152 &\bf{ 1y3y }& 223 &\bf{~~ 3.7 }& 111.9 & 0.10 & 2.40.10.10 & b.47.1.2 \\
 153 &\bf{ 1xi0 }& 143 &\bf{~~ 3.7 }&  41.1 & 0.01 & 2.60.270.20 &   \\
 154 &\bf{ 1h9i }& 223 &\bf{~~ 3.7 }& 110.4 & 0.09 & 2.40.10.10 & b.47.1.2 \\
 155 &\bf{ 2dsc }& 195 &\bf{~~ 3.7 }& 429.0 & 0.56 &   &   \\
 156 &\bf{ 1w4o }& 124 &\bf{~~ 3.7 }& 203.3 & 0.39 & 3.10.130.10 & d.5.1.1 \\
 157 &\bf{ 1lqe }& 223 &\bf{~~ 3.7 }& 114.8 & 0.10 & 2.40.10.10 & b.47.1.2 \\
 158 &\bf{ 1tgn }& 222 &\bf{~~ 3.7 }& 112.6 & 0.09 & 2.40.10.10 & b.47.1.2 \\
 159 &\bf{ 1tnl }& 223 &\bf{~~ 3.7 }& 114.3 & 0.10 & 2.40.10.10 & b.47.1.2 \\
 160 &\bf{ 1otx }& 236 &\bf{~~ 3.6 }& 463.5 & 0.50 & 3.40.50.1580 & c.56.2.1 \\
 161 &\bf{ 1ffe }& 197 &\bf{~~ 3.6 }& 420.9 & 0.55 & 3.40.50.1820 & c.69.1.30 \\
 162 &\bf{ 1bju }& 223 &\bf{~~ 3.6 }& 113.8 & 0.10 & 2.40.10.10 & b.47.1.2 \\
 163 &\bf{ 1anb }& 223 &\bf{~~ 3.6 }& 111.5 & 0.10 & 2.40.10.10 & b.47.1.2 \\
 164 &\bf{ 1ssa }& 113 &\bf{~~ 3.6 }& 163.2 & 0.36 & 3.10.130.10 & d.5.1.1 \\
 165 &\bf{ 1c9p }& 222 &\bf{~~ 3.6 }& 114.1 & 0.10 & 2.40.10.10 & b.47.1.2 \\
 166 &\bf{ 1tx6 }& 223 &\bf{~~ 3.6 }& 111.7 & 0.10 & 2.40.10.10 & b.47.1.2 \\
 167 &\bf{ 2fws }& 139 &\bf{~~ 3.6 }&  91.9 & 0.12 &   & b.1.27.1 \\
 168 &\bf{ 1j16 }& 223 &\bf{~~ 3.6 }& 111.5 & 0.10 & 2.40.10.10 & b.47.1.2 \\
 169 &\bf{ 2g4w }& 124 &\bf{~~ 3.6 }& 204.0 & 0.40 & 3.10.130.10 & d.5.1.1 \\
 170 &\bf{ 3pca }& 200 &\bf{~~ 3.6 }& 251.3 & 0.31 & 2.60.130.10 & b.3.6.1 \\
 171 &\bf{ 3pce }& 200 &\bf{~~ 3.6 }& 251.8 & 0.31 & 2.60.130.10 & b.3.6.1 \\
 172 &\bf{ 1fy8 }& 215 &\bf{~~ 3.6 }& 112.8 & 0.09 & 2.40.10.10 & b.47.1.2 \\
 173 &\bf{ 3pci }& 200 &\bf{~~ 3.6 }& 251.5 & 0.31 & 2.60.130.10 & b.3.6.1 \\
 174 &\bf{ 1vc8 }& 126 &\bf{~~ 3.6 }& 199.1 & 0.37 &   & d.113.1.1 \\
 175 &\bf{ 2a2g }& 158 &\bf{~~ 3.6 }& 208.4 & 0.32 & 2.40.128.20 & b.60.1.1 \\
 176 &\bf{ 2p78 }& 171 &\bf{~~ 3.6 }& 168.7 & 0.23 & 3.40.50.1240 &   \\
 177 &\bf{ 1c78 }& 130 &\bf{~~ 3.6 }& 109.8 & 0.18 & 3.10.20.130 & d.15.5.1 \\
 178 &\bf{ 1xzg }& 197 &\bf{~~ 3.6 }& 421.8 & 0.55 & 3.40.50.1820 & c.69.1.30 \\
 179 &\bf{ 2boc }& 219 &\bf{~~ 3.6 }& 228.3 & 0.23 & 2.60.40.10 & f.14.1.1 \\
 180 &\bf{ 1cuy }& 197 &\bf{~~ 3.6 }& 420.3 & 0.55 & 3.40.50.1820 & c.69.1.30 \\
 181 &\bf{ 2d3j }& 157 &\bf{~~ 3.6 }&  99.8 & 0.07 &   &   \\
 182 &\bf{ 2pqx }& 245 &\bf{~~ 3.6 }& 147.4 & 0.12 &   &   \\
 183 &\bf{ 1ql9 }& 223 &\bf{~~ 3.6 }& 110.7 & 0.10 & 2.40.10.10 & b.47.1.2 \\
 184 &\bf{ 1ntp }& 223 &\bf{~~ 3.6 }& 114.0 & 0.10 & 2.40.10.10 & b.47.1.2 \\
 185 &\bf{ 1fmg }& 223 &\bf{~~ 3.6 }& 115.1 & 0.10 & 2.40.10.10 & b.47.1.2 \\
 186 &\bf{ 1sxt }& 224 &\bf{~~ 3.6 }& 415.9 & 0.48 & 2.40.50.110 & b.40.2.2 \\
 187 &\bf{ 1c2d }& 223 &\bf{~~ 3.6 }& 133.6 & 0.12 & 2.40.10.10 & b.47.1.2 \\
 188 &\bf{ 1ppe }& 223 &\bf{~~ 3.6 }& 113.9 & 0.10 & 2.40.10.10 & b.47.1.2 \\
 189 &\bf{ 1ane }& 223 &\bf{~~ 3.6 }& 113.3 & 0.10 & 2.40.10.10 & b.47.1.2 \\
 190 &\bf{ 1xzb }& 197 &\bf{~~ 3.6 }& 421.2 & 0.55 & 3.40.50.1820 & c.69.1.30 \\

\end{xtabular}
\end{small}
\end{centering}

\vspace*{1cm}
Table 1S.  Continuation of Table 1 of the main text.

\newpage

\clearpage
TABLE 2S: Identification of a mechanical clamp $F_{max}$ for selected proteins. $F_{max}$
denotes the mechanical resistance obtained when all native contacts are present.
$F'_{max}$ is the force obtained when some of some sets of the
relevant native contacts is removed.\\
\vspace*{0.5cm}
\begin{centering}
\begin{small}
\xentrystretch{-0.60}
\begin{xtabular}{c c c c c  } \hline
rank  & PDB  &  $F_{max}$ [$\epsilon$/{\AA}]  & $F'_{max}$ [$\epsilon$/{\AA}]  & $F'_{max}$ [$\epsilon$/{\AA}]           \\ \hline
1  & 1vpf &   5.31                             &  4.72 - slipknot loop            & 1.96   - polymer                    \\
7  & 2h64 &   4.62                             &  4.65 - slipknot loop            & 2.84   - polymer                    \\
19 & 2c7w &   4.23                             &  4.25 - slipknot loop            & 2.15   - polymer                    \\
\hline
\hline
\end{xtabular}
\end{small}
\end{centering}

\newpage

\begin{figure}[!ht]
\epsfxsize=5in
\centerline{\epsffile{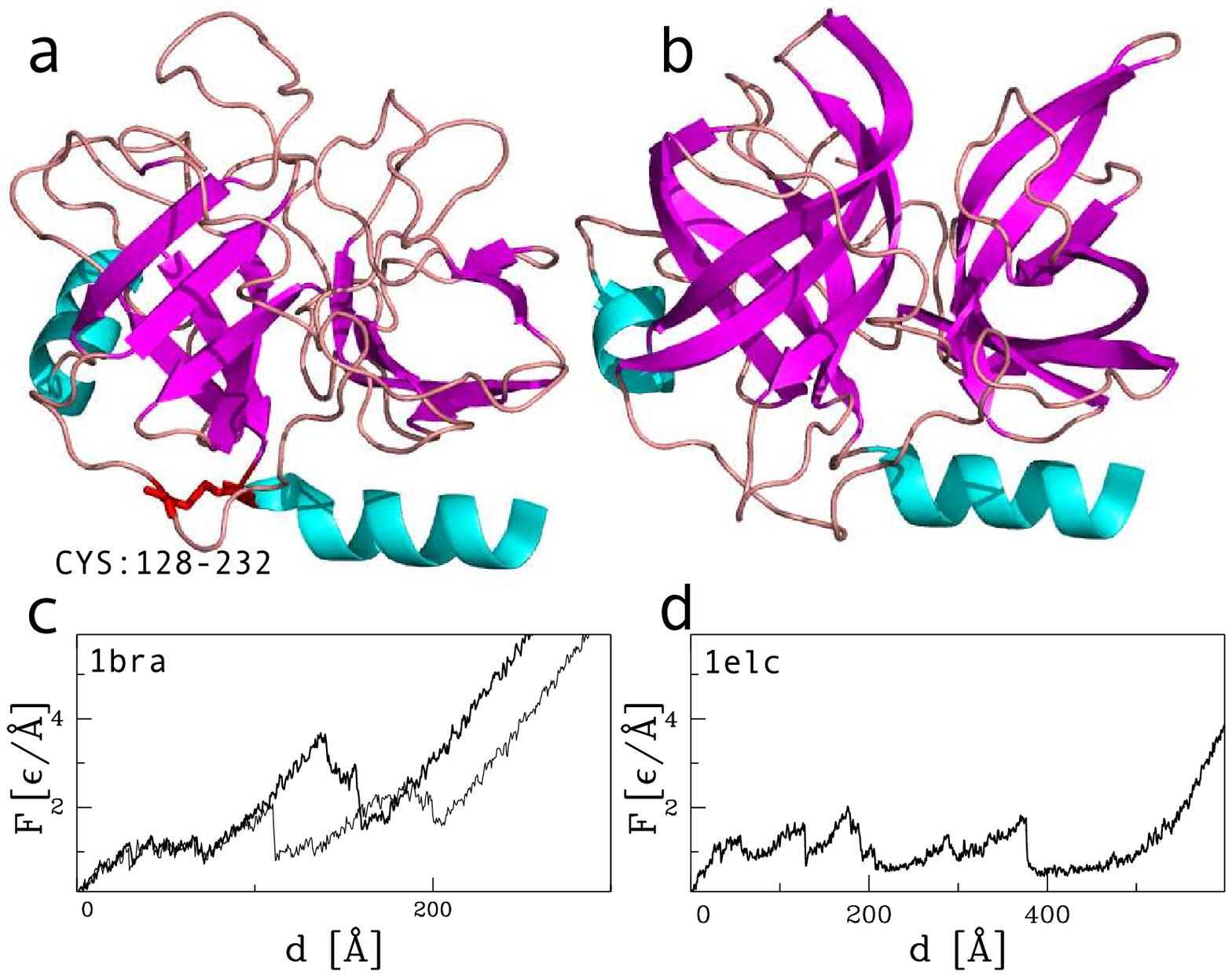}}
\vspace*{-6cm}
\caption*{{\bf Figure S1. Origin of the bimodality in the force distribution for the
SMAD/FHA b.47 fold.} (a) Structure of trypsin 1bra ($N=245$). The mechanically
crucial disulphide bond between sites 128 and 232 is highlighted in red.
(b) Structure of elastase 1elc ($N=255$) which belongs to the same fold b47.1.2 as 1bra.
This structure does not contain two disulphide bonds that 1bra does.
(c) The force-displacement plot for 1bra. $F_{max}$ corresponds to 3.7 $\epsilon$/{\AA}.
The thinner line is obtained when the 128-232 disulphide bond is eliminated -- $F_{max}$
drops to 2.7 $\epsilon$/{\AA}. When one more disulphide bond is cut, stretching
continues to distances shown in panel (d) without affecting $F_{max}$.
(d) The force-displacement plot for 1elc. The
corresponding $F_{max}$ is 2.0 $\epsilon$/{\AA}.
In the case of 1elc, stretching results in the terminal
helix pulling $\beta$ strands from the inside of the protein and thus causing
the inner $\beta$-barrel to unfold. If the case of 1bra (with the disulphide
bridge), the terminal helix pulls the neighbouring loop.
After this event, resistance grows linearly and forms one major force peak.
After the peak, the whole structure opens suddenly,
rupturing contacts between strands in the $\beta$-barrel and in the neighbouring loops.
}
\end{figure}

\end{document}